\documentclass[format=acmsmall,screen]{acmart}

\AtBeginDocument{%
  \providecommand\BibTeX{{%
    \normalfont B\kern-0.5em{\scshape i\kern-0.25em b}\kern-0.8em\TeX}}}

\usepackage{bbm}
\usepackage{todonotes}

\usepackage{graphicx}  
\usepackage{booktabs} 
\usepackage{caption}
\usepackage{subcaption}
\usepackage{enumerate}
\usepackage{multirow}

\newcommand{\algname}[1] {{\fontfamily{cmtt}\selectfont {#1}}}
\definecolor{darkgreen}{rgb}{0.0, 0.5, 0.0}

\usepackage{dsfont}

\usepackage{mathtools}
\DeclarePairedDelimiter{\ceil}{\lceil}{\rceil}

\usepackage[english]{babel}
\usepackage[utf8]{inputenc}
\usepackage{amsmath}
\usepackage{amsfonts}
\usepackage{graphicx}
\usepackage{algorithm}
\usepackage{algpseudocode}
\usepackage{siunitx}
\usepackage{tikz}

\usepackage{xcolor}

\usepackage{geometry}
\begin{document}

\title{A Graph-based Approach for Mitigating Multi-sided Exposure Bias in Recommender Systems}

\author{Masoud Mansoury}
\authornote{This author also has affiliation in School of Computing, DePaul University, Chicago, USA, mmansou4@depaul.edu.}
\affiliation{%
  \institution{Eindhoven University of Technology}
  \city{Eindhoven}
  \country{Netherlands}}
\email{m.mansoury@tue.nl}

\author{Himan Abdollahpouri}
\affiliation{
    \institution{Northwestern University}
    \city{Evanston}
    \state{USA}
}
\email{}

\author{Mykola Pechenizkiy}
\affiliation{%
 \institution{Eindhoven University of Technology}
 \city{Eindhoven}
 \country{the Netherlands}}
\email{m.pechenizkiy@tue.nl}

\author{Bamshad Mobasher}
\affiliation{%
  \institution{DePaul University}
  \city{Chicago}
  \country{USA}}
\email{mobasher@cs.depaul.edu}

\author{Robin Burke}
\affiliation{%
  \institution{University of Colorado Boulder}
  \city{Boulder}
  \country{USA}}
\email{robin.burke@colorado.edu}

\begin{abstract}

 Fairness is a critical system-level objective in recommender systems that has been the subject of extensive recent research. A specific form of fairness is supplier exposure fairness where the objective is to ensure equitable coverage of items across all suppliers in recommendations provided to users. This is especially important in multistakeholder recommendation scenarios where it may be important to optimize utilities not just for the end-user, but also for other stakeholders such as item sellers or producers who desire a fair representation of their items. This type of supplier fairness is sometimes accomplished by attempting to increasing aggregate diversity in order to mitigate popularity bias and to improve the coverage of long-tail items in recommendations. In this paper, we introduce FairMatch, a general graph-based algorithm that works as a post processing approach after recommendation generation to improve exposure fairness for items and suppliers. The algorithm iteratively adds high quality items that have low visibility or items from suppliers with low exposure to the users' final recommendation lists. A comprehensive set of experiments on two datasets and comparison with state-of-the-art baselines show that FairMatch, while significantly improves exposure fairness and aggregate diversity, maintains an acceptable level of relevance of the recommendations.
\end{abstract}



\maketitle

\section{Introduction}

Recommender systems use historical data on interactions between users and items to generate personalized recommendations for the users. These systems are used in a variety of different applications including movies, music, e-commerce, online dating, and many other areas where the number of options from which the user needs to choose can be overwhelming. There are many different metrics to evaluate the performance of the recommender systems ranging from accuracy metrics such as precision, normalized discounted cumulative gain (NDCG), and recall to non-accuracy ones like novelty and serendipity \cite{kaminskas2016,vargas2011rank,castells2015novelty}. 

All these metrics mentioned above are usually calculated using each individual recommendation list to a given user. However, there are certain aspects of the recommendations that need to be looked at in a more holistic way across all users. For example, how successful a given recommender system is in giving enough chance to different items to be seen by different users cannot be measured by looking at individual recommendations separately. This is particularly important in domains where fairness of the recommendations can be critical for the success of the recommendation platform such as job recommendation, philanthropy, online dating, etc. 

One of the main reasons for different items not getting a fair exposure in the recommendations is the popularity bias problem where few popular items are over-recommended while the majority of other items do not get a deserved attention. For example, in a music recommendation system, few popular artists might take up the majority of the streamings leading to under-exposure of less popular artists. This bias, if not mitigated, can negatively affect the experience of different users and items on the platform \cite{mehrotra2018towards,abdollahpouri2020multi}. It could also be perpetuated over time by the interaction of users with biased recommendations and, as a result, using biased interactions for training the model in the subsequent times \cite{damour2020,chaney2018,sinha2016,sun2019debiasing,mansoury2020feedback}. 

There are numerous methods to tackle this popularity bias by either modifying the underlying recommendation algorithms by incorporating the popularity of each item 
\cite{vargas2014improving,sun2019debiasing,abdollahpouri2017controlling,adamopoulos2014over} or as a post-processing re-ranking step to modify an existing, often larger, recommendation list and extract a shorter list that has a better characteristics in terms of fair exposure of different items \cite{adomavicius2011maximizing,adomavicius2011improving,antikacioglu2017,abdollahpouri2019managing}. However, most of these of algorithms solely concentrated on mitigating the exposure (visibility) bias in an item level. What these algorithms ignore is the complexity of many real world recommender systems where there are different suppliers that provide the recommended items and hence the fairness of exposure in a supplier level need to be also addressed \cite{himan2020a}. One way to improve supplier exposure fairness is to improve the visibility of items hoping it will also lead to giving a more balanced exposure to different suppliers as often there is a positive correlation between the popularity of suppliers and their items. However, only optimizing for item visibility without explicitly taking into account the suppliers in the recommendations does not necessarily make the recommendations fairer for suppliers as we have explained in Section \ref{section3}. 

In this paper, we introduce \textit{FairMatch}, a general graph-based algorithm that works as a post-processing approach after recommendation generation (on top of any existing standard recommendation algorithm). The idea is to generate a list of recommendations with a size larger than what we ultimately want for the final list using a standard recommendation algorithm and then use our FairMatch algorithm to build the final list using a subset of items in the original list. This is done by iteratively solving \textit{Maximum Flow} problem on a recommendation bipartite graph which is built using the recommendations in the original list (left nodes are recommended items and right nodes are the users). At each iteration, the items that can be good candidates for the final list will be selected and removed from the graph and the process will continue on the remaining part of the graph. FairMatch is able to systematically improve the visibility of different items and suppliers in the recommendations while maintaining an acceptable level of relevance to the users. This work is an extension of our previously proposed algorithm that was only capable of improving item visibility \cite{fairmatch2020umap}. This extended version, on the other hand, is able to improve the visibility of items or suppliers of the items depending on which one we pick as our optimization criterion. 

To show the effectiveness of our FairMatch algorithm on improving aggregate diversity and fair visibility of recommended items and suppliers, we perform a comprehensive set of experiments using three standard recommendation algorithms as the base for our reranking method. Comparison with several state-of-the-art baselines shows that our FairMatch algorithm is able to significantly improve the performance of recommendation results in terms of visibility of different items and suppliers with a negligible loss in the recommendation accuracy in some cases. 

In summary, we make the following contributions:
\begin{itemize}
    \item We propose a graph-based approach for improving the exposure fairness of items and suppliers in the recommendations. 
    \item We propose several metrics that can better measure the effectiveness of an algorithm in mitigating the exposure bias for items and suppliers.
    \item We compare our proposed algorithm with several state-of-the-art methods in mitigating popularity and exposure bias and show the effectiveness of our algorithm in doing so. 
\end{itemize}

\section{Related Work}

Research on fairness and algorithmic bias have revealed that there are different types of bias in recommendation systems that, if not mitigated, can negatively affect the system \cite{chen2020bias}. Various algorithmic solutions have been proposed to address this issue. For example, Yao and Huang in \cite{yao2017beyond} improved the fairness of recommendation results by adding fairness terms to objective function in model-based recommendation algorithms. As another example, Liu et. al. in \cite{liu2020balancing} proposed a reinforcement learning based framework to address the trade-off between accuracy and fairness in interactive recommender systems. These works mainly addressed group fairness when users are grouped based on a sensitive attribute. There are also studies addressing other types of bias such as position bias \cite{biega2018equity}. However, this paper focuses on exposure bias in recommendation results. In this section, we review the literature on exposure bias in recommender systems. 

The concept of popularity bias has been studied by many researchers often under different names such as long-tail recommendation \cite{abdollahpouri2017controlling,yin2012challenging}, Matthew effect \cite{moller2018not}, and aggregate diversity \cite{liu2015trust,adomavicius2011improving} all of which refer to the fact that the recommender system should recommend a wider variety of items across all users. 

The solutions for tackling popularity bias in the literature can be categorized into two groups:\footnote{A third option, preprocessing, is generally not useful for popularity bias mitigation because undersampling the popular items greatly increases the sparsity of the data.} \textit{Model-based} \cite{vargas2014improving,sun2019debiasing,abdollahpouri2017controlling,adamopoulos2014over} and \textit{re-ranking} \cite{adomavicius2011maximizing,adomavicius2011improving,antikacioglu2017,abdollahpouri2019managing}. In model-based solutions the recommendation generation step is modified, so that the popularity of the items is taken into account in the rating prediction. For example authors in \cite{abdollahpouri2017controlling} proposed a regularization term to control the popularity of recommended items that could be added to an existing objective function of a learning-to-rank algorithm \cite{karatzoglou2013learning} to improve the aggregate diversity of the recommendations. In another work, Vargas and Castells in \cite{vargas2011rank} proposed probabilistic models for improving novelty and diversity of recommendations by taking into account both relevance and novelty of target items when generating recommendation lists. Moreover, authors in  \cite{vargas2014improving}, proposed the idea of recommending users to items for improving novelty and aggregate diversity. They applied this idea to nearest neighbor models as an inverted neighbor and a factorization model as a probabilistic reformulation that isolates the popularity components. 

In the second approach for improving aggregate diversity, the algorithm takes a larger output recommendation list and re-orders the items in the list to extract a shorter final list with improved long-tail properties. Most of the solutions for tackling popularity bias fall into this category. For example, Adomavicius and Kwon \cite{adomavicius2011maximizing} proposed the idea of diversity maximization using a maximum flow approach. They used a specific setting for the bipartite recommendation graph in a way that the maximum amount of flow that can be sent from a source node to a sink node would be equal to the maximum aggregate diversity for those recommendation lists. In their setting, given the number of users is $m$, the source node can send a flow of up to $m$ to the left nodes, left nodes can send a flow of up to 1 to the right nodes, and right nodes can send a flow of up to 1 to the sink node. Since the capacity of left nodes to right nodes is set to 1, thus the maximum possible amount of flow through that recommendation bipartite graph would be equivalent to the maximum aggregate diversity. 

A more recent graph-based approach for improving aggregate diversity which also falls into the reranking category was proposed by Antikacioglu and Ravi in \cite{antikacioglu2017}. They generalized the idea proposed in \cite{adomavicius2011maximizing} and showed that the minimum-cost network flow method can be efficiently used for finding recommendation subgraphs that optimizes the diversity. In this work, an integer-valued constraint and an objective function are introduced for discrepancy minimization. The constraint defines the maximum number of times that each item should appear in the recommendation lists and the objective function aims to find an optimal subgraph that gives the minimum discrepancy from the constraint. This work shows improvement in aggregate diversity of the items with a smaller accuracy loss compared to the work in \cite{vargas2011rank} and \cite{vargas2014improving}. Our algorithm is also a graph-based approach that not only is it able to improve aggregate diversity and the exposure fairness of items, it also gives the suppliers of the recommended items a fairer chance to be seen by different users. Moreover, unlike the work in \cite{antikacioglu2017} which tries to minimize the discrepancy between the distribution of the recommended items and a target distribution, our FairMatch algorithm has more freedom in promoting high-quality items or suppliers with low visibility since it does not assume any target distribution of the recommendation frequency.

Another work that also uses a re-ranking approach is by Abdollahpouri et al. \cite{abdollahpouri2019managing} where authors proposed a diversification method for improving aggregate diversity and long-tail coverage in recommender systems. Their method was based on \textit{eXplicit Query Aspect Diversification} (xQuAD) algorithm \cite{santos2010exploiting} that was designed for diversifying the query result such that it covers different aspects related to a given query. In \cite{abdollahpouri2019managing}, the authors used xQuAD algorithm for balancing the ratio of popular and less popular (long-tail) items in final recommendation lists. 

In addition, in \cite{fairmatch2020umap}, we proposed a graph-based algorithm that finds high quality items that have low visibility in the recommendation lists by iteratively solving the maximum flow problem on recommendation bipartite graph. The present work extends the idea in \cite{fairmatch2020umap} for improving the exposure fairness of items and suppliers. We also discuss the limitations of existing metrics for measuring the exposure bias in recommendation results and propose several metrics to overcome those limitations. 

In addressing exposure bias in domains like job recommendations where job seekers or qualified candidates are recommended, Zehlike et. al. \cite{zehlike2017fa} proposed a re-ranking algorithm to improve the ranked group fairness in recommendations. The algorithm creates queues of protected and unprotected items and merges them using normalized scoring such that protected items get more exposure. Singh and Joachims in \cite{singh2018fairness} discussed how exposure bias can lead to unfair treatment of different groups of users in ranking systems. They proposed a general framework for addressing the exposure bias when maximizing the utility for users in generating the ranked results. Geyik et. al. in \cite{geyik2019fairness} explored the exposure bias in LinkedIn Talent Search where the distribution of applicants belong to different groups of sensitive attributes in recommendation lists do not follow the distribution of applicants' group in the initial search results. They showed that applicants belong to the protected group are often under-recommended. To address this issue, the authors proposed an algorithm to achieve the desired distribution of applicants' group with respect to sensitive attributes in topn recommendation results.  

Most of the existing works in the literature for improving aggregate diversity and exposure fairness have only concentrated on the items and ignored the fact that in many recommendation domains the recommended items are often provided by different suppliers and hence their utility should also be investigated. To the best of our knowledge, there are only few prior works that have addressed this issue such as \cite{abdollahpouri2020addressing} and \cite{mehrotra2018towards}. In \cite{abdollahpouri2020addressing}, authors illustrated how popularity bias is a multistakeholder problem and hence they evaluated their solution for mitigating this bias from the perspective of different stakeholders. Mehrotra et al. \cite{mehrotra2018towards} investigated the trade-off between the relevance of recommendations for users and supplier fairness, and their impacts on users' satisfaction. Relevance of the recommended items to a user is determined by the score predicted by a recommendation algorithm. To determine the supplier fairness in recommendation list, first, suppliers are grouped into several bins based on their popularity in rating data and then the supplier fairness of a recommendation list is measured as how diverse the list is in terms of covering different supplier popularity bins. 

Our work in this paper also observes the importance of evaluating algorithms from the perspective of multiple stakeholders and we propose an algorithm that can directly improve the visibility of the suppliers without losing much accuracy from the users' perspective.  

\section{Exposure Bias in Recommendation}\label{section3}

It is well-known that recommendation algorithms favor popular items which leads to an unfair exposure of other items that might not be as popular \cite{harald2011,abdollahpouri2017controlling}. This bias towards popular items can negatively affect the less popular items, items that are new to the system (aka cold start items), and even the supplier of the items \cite{himan2020a,patro2020fairrec}. In this section, we illustrate the exposure bias of several recommendation algorithms from both the items and suppliers perspective. The mathematical and formulated definitions of exposure for items or suppliers are presented in section \ref{metrics}. 

\begin{figure*}[btp]
    \centering
    \begin{subfigure}[b]{0.33\textwidth}
        \includegraphics[width=\textwidth]{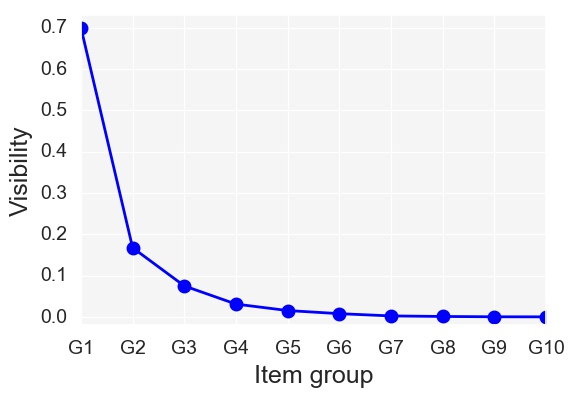}
        \caption{BPR} \label{dist_ml_50_item_bpr}
    \end{subfigure}
    \begin{subfigure}[b]{0.33\textwidth}
        \includegraphics[width=\textwidth]{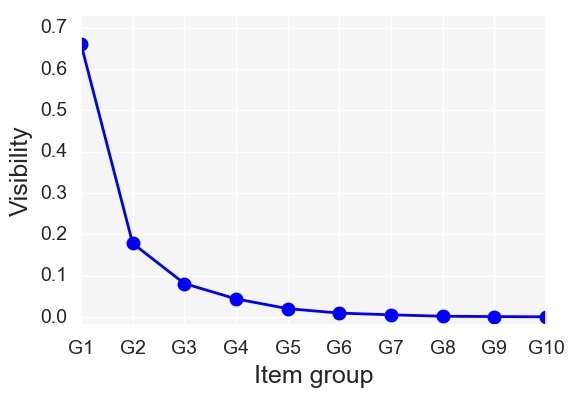}
        \caption{NCF} \label{dist_ml_50_item_ncf}
    \end{subfigure}%
    \begin{subfigure}[b]{0.33\textwidth}
        \includegraphics[width=\textwidth]{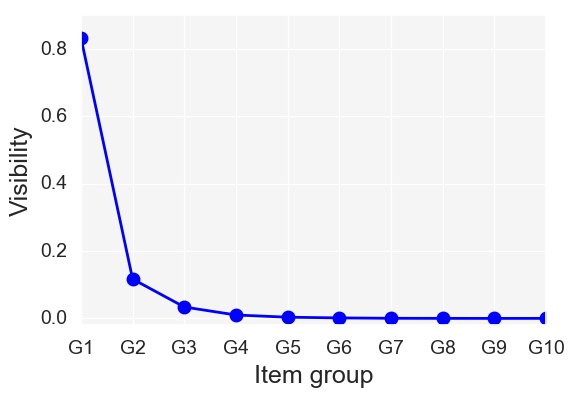}
        \caption{UserKNN} \label{dist_ml_50_item_userknn}
    \end{subfigure}%
\caption{Visibility of recommended items for different recommendation algorithms on MovieLens dataset.} \label{dist_ml_50_item}
\end{figure*}

\begin{figure*}[btp]
    \centering
    \begin{subfigure}[b]{0.33\textwidth}
        \includegraphics[width=\textwidth]{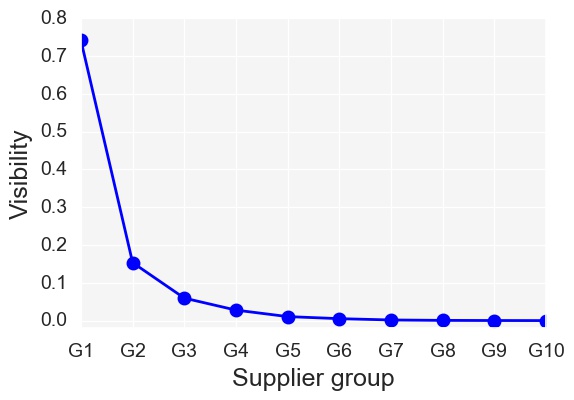}
        \caption{BPR} \label{dist_ml_50_sup_bpr}
    \end{subfigure}
    \begin{subfigure}[b]{0.33\textwidth}
        \includegraphics[width=\textwidth]{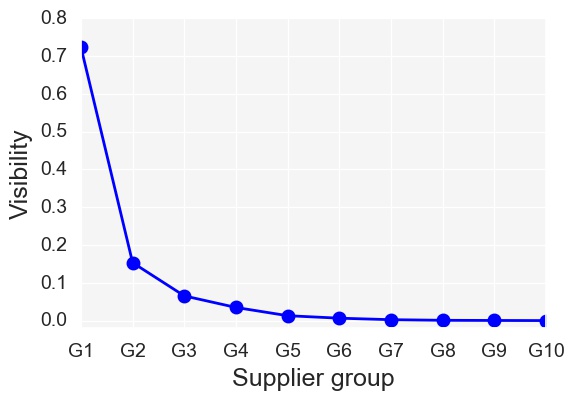}
        \caption{NCF} \label{dist_ml_50_sup_ncf}
    \end{subfigure}%
    \begin{subfigure}[b]{0.33\textwidth}
        \includegraphics[width=\textwidth]{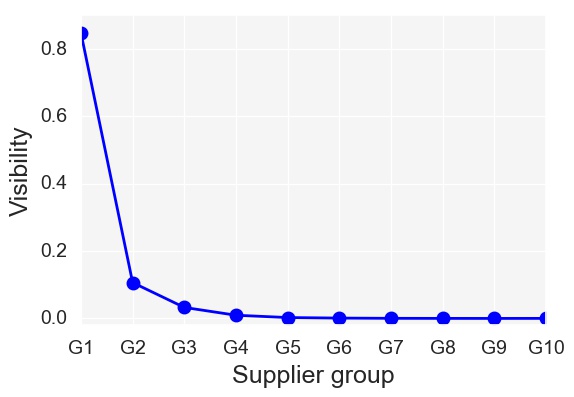}
        \caption{UserKNN} \label{dist_ml_50_sup_userknn}
    \end{subfigure}%
\caption{Visibility of suppliers for different recommendation algorithms on MovieLens dataset.} \label{dist_ml_50_sup}
\end{figure*}

\subsection{Bias in Item Exposure}

An exposure for an item is the percentage of the times it has appeared in the recommendations \cite{khenissi2020modeling,singh2018fairness}. Recommendation algorithms are often biased towards more popular items giving them more exposure than many other items. Figure \ref{dist_ml_50_item} shows the visibility of different items in the recommendations produced by three recommendation algorithms: Bayesian Personalized Ranking (\algname{BPR}) \cite{rendle2009bpr}, Neural Collaborative Filtering (\algname{NCF}) \cite{he2017neural}, and User-based Collaborative Filtering (\algname{UserKNN}) \cite{Resnick:1994a}. Items are binned into ten equal-size groups based on their visibility in recommendation lists such that each bin contains the same number of items. In other words, items are first sorted based on their visibility in recommendation lists and then ten equal-size groups of items are created. We can see that in all three algorithms, there is a long-tail shape for the visibility of the items indicating few popular item groups are recommended much more frequently than the others creating an item exposure bias in the recommendations. Not every algorithm has the same level of exposure bias for different items. For instance, we can see that \algname{UserKNN} has recommended items from group $G_1$ to roughly 80\% of the users while this number is near 70\% and 65\% for \algname{BPR} and \algname{NCF}, respectively. On the other hand, $G_2$ has received less exposure in \algname{UserKNN} (10\%) compared to \algname{BPR} and \algname{NCF} which have given 17\% and 19\% visibility to items in this group, respectively. 

\begin{figure}[btp]
    \centering
    \includegraphics[width=0.5\textwidth]{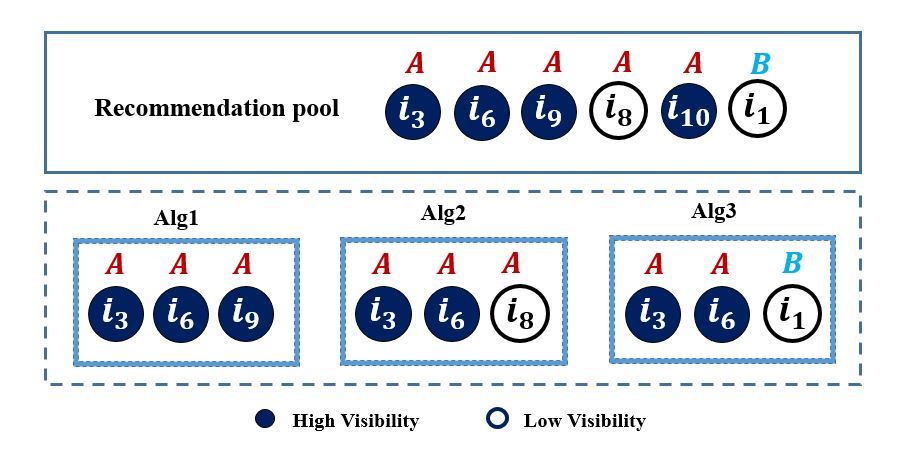}
    \caption{Comparison between a relevance based recommendation algorithm ($Alg1$), item visibility-aware reranker ($Alg2$), and supplier visibility-aware reranker ($Alg3$).}\label{motivation}
\end{figure}

\subsection{Bias in Supplier Exposure}

The unfair exposure does not only affect the items in a recommender system. We know that in many recommendation platforms the items to be recommended are provided by different suppliers. Therefore, the dynamic of how recommendation algorithms can impact the experience of the suppliers is also crucial. Authors in \cite{abdollahpouri2020multi} empirically show that recommendation algorithms often over-promote items from popular suppliers while suppressing the less popular ones. Figure \ref{dist_ml_50_sup} shows a similar plot to Figure \ref{dist_ml_50_item} but for the suppliers of the items. Similar to items, suppliers are binned into ten groups based on their visibility in recommendation lists. The same problem that we observed in Figure \ref{dist_ml_50_item} also exists here: in all three algorithms, there is a long-tail shape for the visibility of the suppliers indicating few supplier groups are recommended much more frequently than the others. 

There are many existing works for improving the visibility of different items in the recommendations and reducing the exposure bias in items. However, the same cannot be said about the suppliers and there has not been much attention to improving the supplier visibility/exposure. Although, improving the item visibility can, indirectly, help suppliers as well as it was demonstrated in \cite{abdollahpouri2020addressing}, a more explicit incorporation of suppliers in the recommendation process can yield fairer outcomes for different suppliers in terms of visibility. 

Our work in this paper aims to address this problem by directly incorporating the suppliers in the recommendation process to mitigate the exposure bias from the suppliers perspective.

\subsection{Motivating Example}

Figure \ref{motivation} shows a scenario where we have a list of items as candidate pool and the goal is to extract a list of recommendations (in this example the size is 3 for illustration purposes) and recommend it to the user. In addition, items are categorized to either \textit{high visibility} (i.e. frequently recommended) and \textit{low visibility} (less frequently recommended). Moreover, each item is also provided by either supplier $A$ or $B$. Three recommendation algorithms (these are just for illustration purposes) are compared in terms of how they extract the final list of three items. The first algorithm \textit{Alg1} extracts the three most relevant items from the top of the list without considering the visibility of items or which supplier they belong to. Obviously, this algorithm performs poorly in terms the fairness of item exposure and supplier exposure since only highly relevant items are recommended and they are all from supplier $A$. In contrast, the second algorithm \textit{Alg2} extracts the final recommendation list by also taking into account the visibility of items. This algorithm could represent many existing approaches to overcome exposure bias in recommendation. However, although the list of recommended items are now more diverse in terms of different type of items (high visibility vs low visibility) it still only contains items from supplier $A$ since the supplier information was not incorporated in the algorithm. The third algorithm \textit{Alg3}, on the other hand, has recommended a diverse list of items not only in terms of items, but also in terms of the suppliers of those items. This is the type of algorithm we intend to develop in this paper. 

In the next section, we present our algorithm, FairMatch, which can systematically improve the exposure of less popular items and suppliers while maintaining an acceptable level of relevance to the users. 

\section{FairMatch Algorithm}

We formulate our FairMatch algorithm as a post-processing step after the recommendation generation. In other words, we first generate recommendation lists of larger size than what we ultimately desire for each user using any standard recommendation algorithm and use them to build the final recommendation lists. FairMatch works as a batch process, similar to that proposed in \cite{surer2018} where all the recommendation lists are produced at once and re-ranked simultaneously to achieve the objective. In this formulation, we produce a longer recommendation list of size $t$ for each user and then, after identifying candidate items (based on defined utility, more details in section \ref{weight_comp}) by iteratively solving the maximum flow problem on recommendation bipartite graph, 
we generate a shorter recommendation list of size $n$ (where $t>>n$).

Let $G=(I,U,E)$ be a bipartite graph of recommendation lists where $I$ is the set of left nodes, $U$ is the set of right nodes, and $E$ is the set of edges between left and right nodes when recommendation occurred. $G$ is initially a uniformly weighted graph, but we will update the weights for edges as part of our algorithm. We will discuss the initialization and our weighting method in section \ref{weight_comp}. 

Given a weighted bipartite graph $G$, the goal of our FairMatch algorithm is to improve the exposure fairness of recommendations without a significant loss in accuracy of the recommendations. We define exposure fairness as providing equal chance for items or suppliers to appear in recommendation lists. The FairMatch algorithm does this by identifying items or suppliers with low visibility in recommendation lists and promote them in the final recommendation lists while maintaining the relevance of recommended items for users. We develop our algorithm by extending the approach introduced in \cite{garcia2020fair} to improve the exposure fairness of the recommender systems.

We use an iterative process to identify the subgraphs of $G$ that satisfy the underlying definitions of fairness without a significant loss in accuracy of the recommendation for each user. After identifying a subgraph $\Gamma$ at each iteration, we remove $\Gamma$ from $G$ and continue the process of finding subgraphs on the rest of the graph (i.e., $G / \Gamma$). We keep track of all the subgraphs as we use them to generate the final recommendations in the last step. 

Identifying $\Gamma$ at each iteration is done by solving a \textit{Maximum Flow} problem (explained in section \ref{fair_max}) on the graph obtained from the previous iteration. Solving the maximum flow problem returns the left nodes connected to the edges with lower weight on the graph. After finding those left nodes, we form subgraph $\Gamma$ by separating identified left nodes and their connected right nodes from $G$. Finally, $<user,item>$ pairs in subgraphs are used to construct the final recommendation lists of size $n$. We will discuss this process in detail in the following sections.

\begin{algorithm}[btp]
\caption{The FairMatch Algorithm}
\small
\begin{algorithmic}
  \Function{FairMatch}{Recommendations $R$, TopN $n$, Suppliers $S$, Coefficient $\lambda$}
    \State Build graph $G=(I,U,E)$ from $R$
    \State Initialize \textit{subgraphs} to empty
    \Repeat
        \State $G$=WeightComputation($G$, $R$, $S$, $\lambda$)
        \State $\mathcal{I}_{C}$ = Push-relabel($G$)
        \State Initialize $subgraph$ to empty
        \For {\textbf{each} $i \in \mathcal{I}_{C}$}
            \If{$label_{i} \geq |I|+|U|+2$}
                \For {\textbf{each} $u \in Neighbors(i)$}
                    \State Append $<i,u,e_{iu}>$ to \textit{subgraph}
                \EndFor
            \EndIf 
        \EndFor
        \If{$subgraph$ is empty}
            \State $break$
        \EndIf
        \State Append $subgraph$ to $subgraphs$
        \State $G$=Remove $subgraph$ from $G$
    \Until($true$)
    \State Reconstruct $R$ of size $n$ based on \textit{subgraphs}
   \EndFunction
\end{algorithmic}

\end{algorithm}

\begin{figure}[btp]
    \centering
    \includegraphics[width=0.99\textwidth]{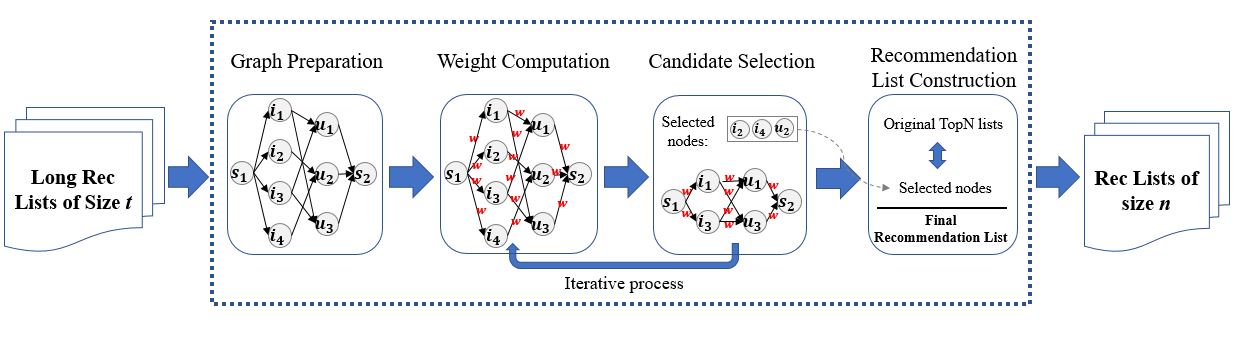}
    \caption{The process of FairMatch algorithm.}\label{fig:process}
\end{figure}

Algorithm 1 shows the pseudocode for FairMatch. Overall, our FairMatch algorithm consists of the following four steps: 1) Graph preparation, 2) Weight computation, 3) Candidate selection, and 4) Recommendation list construction. Figure \ref{fig:process} shows the process of FairMatch algorithm. FairMatch takes the long recommendation lists of size $t$ generated by a base recommendation algorithm as input, and then over four consecutive steps, as mentioned above, it generates the final recommendation lists. The detail about each step in FairMatch algorithm would be discussed in the following subsections. 

\subsection{Graph Preparation}

\begin{figure*}[t]
    \centering
    \includegraphics[width=0.3\textwidth]{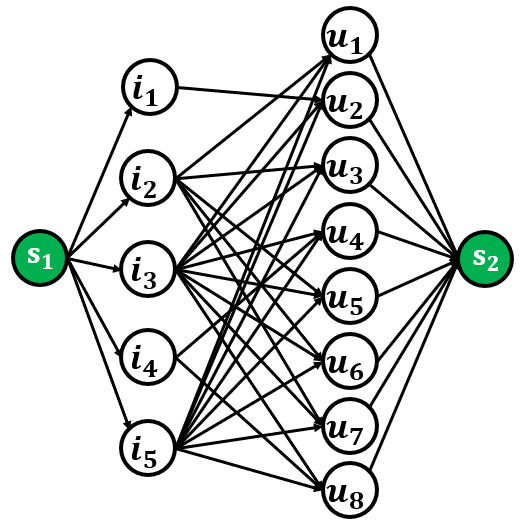}
    \caption{An example of a recommendation bipartite graph of recommendation lists of size 3.}\label{fig:ex}
\end{figure*}

Given long recommendation lists of size $t$ generated by a standard recommendation algorithm, we create a bipartite graph from recommendation lists in which items and users are the nodes and recommendations are expressed as edges. Since our FairMatch algorithm is formulated as a maximum flow problem, we also add two nodes, \textit{source} ($s_1$) and \textit{sink} ($s_2$). The purpose of having a source and sink node in the maximum flow problem is to have a start and endpoint for the flow going through the graph. We connect $s_1$ node to all left nodes and also we connect all right nodes to $s_2$. Figure \ref{fig:ex} shows a sample bipartite graph resulted in this step. 

\subsection{Weight Computation}\label{weight_comp}

Weight computation step plays an important role on improving the exposure fairness of recommendations in the proposed model. Depending on the fairness definition that we desire to achieve, weight computation step should be adapted accordingly. In this subsection, we discuss how weight computation can be adapted for improving the exposure fairness of items or suppliers.

Given the bipartite recommendation graph, $G=(I,U,E)$, the task of weight computation is to calculate the weight for edges between the source node and left nodes, left nodes and right nodes, and right nodes and sink node.

For edges between left nodes and right nodes, we define the weights as the weighted sum of user utility and supplier utility (or instead, item utility). The utility of each user is defined as the relevance of recommended items for that user. Given the long recommendation list of size $t$ for user $u$ as $L_u$, in our formulation, we define the relevance of an item $i$ for user $u$ as rank of $i$ in sorted $L_u$ in descending order based on predicted score by the base recommender. This way, items in lower rank will be more relevant to the user (e.g. item in the first rank is the most relevant one). 

The utility for each item and supplier is defined as their exposure or visibility in the long recommendation lists. The visibility of each item is defined as the degree of the node corresponding to that item (excluding the edge with the source node). Item degree is the number of edges going out from that node connecting it to the user nodes and that shows how often it is recommended to different users. Analogously, the visibility for each supplier is defined as sum of the degree of all nodes corresponding to the items belonging to that supplier. Therefore, we introduce two separate weight computation schemes, one for item utility and another for supplier utility, which eventually results in two variations of FairMatch algorithm as follows:

\begin{itemize}
    \item \textbf{$FairMatch^{item}$:} For computing the weight for edges between $i \in I$ and $u \in U$, we use the following equation:

\begin{equation}\label{wi}
    w_{iu}= \lambda \times rank_{iu} + (1 - \lambda) \times degree_i
\end{equation}

\noindent where $rank_{iu}$ is the position of item $i$ in the sorted recommendation list of size $t$ generated for user $u$, $degree_i$ is the number of edges from $i$ to right nodes (i.e., $u \in U$), and $\lambda$ is a coefficient to control the trade-off between the relevance of the recommendations and the exposure of items. 

    \item \textbf{$FairMatch^{Sup}$}: For computing the weight for edges between $i \in I$ and $u \in U$, we use the following equation:
  
  \begin{equation}\label{ws}
    w_{iu}= \lambda \times rank_{iu} + (1 - \lambda) \times       
    \sum_{i \in A(B(i))}{degree_{i}}
\end{equation}

    where $B(i)$ returns the supplier of item $i$ and $A(B(i))$ returns all items belonging to the supplier of item $i$. Therefore, the term $\sum_{i \in A(B(i))}{degree_{i}}$ computes the visibility of supplier of item $i$ (i.e., sum of visibility of all items that belong to the supplier of item $i$). $rank_{iu}$ and  $\lambda$ have the same definition as Equation \ref{wi}. 
\end{itemize}

Note that in equation \ref{wi} and \ref{ws}, $rank_{iu}$ and visibility for suppliers and items have different ranges. The range for $rank_{iu}$ is from 1 to $t$ (there are $t$ different positions in the original list) and the range of visibility depends on  the frequency
of the item (or its supplier) recommended to the users (the more frequent it is recommended to different users the higher its degree is). Hence, for a meaningful weighted sum, we normalize visibility of items and suppliers to be in the same range as $rank_{iu}$. 

Given weights of the edges between $i \in I$ and $u \in U$, $w_{iu}$, total capacity of $I$ and $U$ would be 
$C_{T}=\sum_{i\in I}^{}\sum_{u \in U}^{}w_{iu}$ which simply shows the sum of the weights of the edges connecting left nodes to the right nodes.

For computing the weight for edges connected to the source and sink nodes, first, we equally distribute $C_{T}$ to left and right nodes. Therefore, the capacity of each left node, $C_{eq}(I)$, and right node, $C_{eq}(U)$, would be as follow: 

\begin{equation}
C_{eq}(I)=\ceil[\bigg]{\frac{C_{T}}{|I|}}, \;\;\;\;\;\; C_{eq}(U)=\ceil[\bigg]{\frac{C_{T}}{|U|}}
\end{equation}

\noindent where $\ceil[\big]{a}$ returns the ceil value of $a$. Then, based on equal capacity assigned to each left and right nodes, we follow the method introduced in \cite{garcia2020fair} to compute weights for edges connected to source and sink nodes as follow:

\begin{equation} \label{eq:sl_cap1}
\forall i \in I, w_{s_{1}i}=\ceil[\bigg]{min(\frac{C_{eq}(I)}{gcd(C_{eq}(I),C_{eq}(U))},\frac{C_{eq}(U)}{gcd(C_{eq}(I),C_{eq}(U))})}
\end{equation}

\begin{equation} \label{eq:sl_cap2}
\forall u \in U, w_{us_{2}}=\ceil[\bigg]{\frac{C_{eq}(I)}{gcd(C_{eq}(I),C_{eq}(U))}}
\end{equation}

\noindent where $gcd(C_{eq}(I),C_{eq}(U))$ is the Greatest Common Divisor of the distributed capacity of left and right nodes. Assigning the same weight to edges connected to the source and sink nodes guaranties that all nodes in $I$ and $U$ are treated equally and the weights between them play an important role in our FairMatch algorithm.

\subsection{Candidate Selection}\label{fair_max}

The graph constructed in previous steps is ready to be used for solving the maximum flow problem. In a maximum flow problem, the main goal is to find the maximum amount of feasible flow that can be sent from the source node to the sink node through the flow network. Several algorithms have been proposed for solving a maximum flow problem. Well-known algorithms are Ford--Fulkerson \cite{ford1956}, Push-relabel \cite{goldberg1988}, and Dinic's algorithm \cite{dinic1970}. In this paper, we use Push-relabel algorithm to solve the maximum flow problem on our bipartite recommendation graph as it is one of the efficient algorithms for this matter and also it provides some functionalities that our FairMatch algorithm benefits them.

\begin{figure*}[t]
    \centering
    \begin{subfigure}[b]{0.32\textwidth}
        \includegraphics[width=\textwidth]{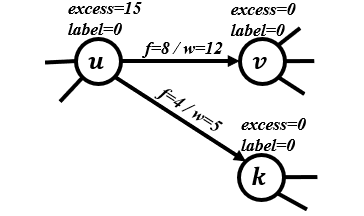}
        \caption{Original graph} \label{fig:push_ex:a}
    \end{subfigure}
    \begin{subfigure}[b]{0.33\textwidth}
        \includegraphics[width=\textwidth]{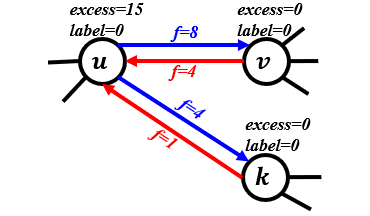}
        \caption{Residual graph} \label{fig:push_ex:b}
    \end{subfigure}%
    \begin{subfigure}[b]{0.33\textwidth}
        \includegraphics[width=\textwidth]{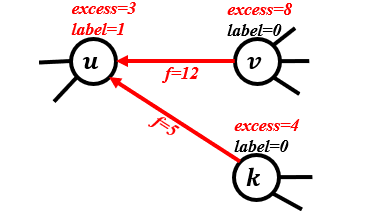}
        \caption{Pushing excess flow of u} \label{fig:push_ex:c}
    \end{subfigure}%
\caption{Example of push and relabel operations.} \label{fig:push_ex}
\end{figure*}

In push-relabel algorithm, each node will be assigned two attributes: \textit{label} and \textit{excess flow}. The label attribute is an integer value that is used to identify the neighbors to which the current node can send flow. A node can only send flow to neighbors that have lower label than the current node. Excess flow is the remaining flow of a node that can still be sent to the neighbors. When all nodes of the graph have excess flow equals to zero, the algorithm will terminate.

The push-relabel algorithm combines $push$ operations that send a specific amount of flow to a neighbor, and $relabel$ operations that change the label of a node under a certain condition (when the node has excess flow greater than zero and there is no neighbor with label lower than the label of this node).

Here is how the push-relabel algorithm works: Figure \ref{fig:push_ex} shows a typical graph in the maximum flow problem and an example of push and relabel operations. In Figure \ref{fig:push_ex:a}, $f$ and $w$ are current flow and weight of the given edge, respectively. In Push-relabel algorithm, a residual graph, $G^{'}$, will be also created from graph $G$. As graph $G$ shows the flow of forward edges, graph $G^{'}$ shows the flow of backward edges calculated as $f_{backward}=w-f$. Figure \ref{fig:push_ex:b} shows residual graph of graph $G$ in Figure \ref{fig:push_ex:a}. Now, we want to perform a push operation on node $u$ and send its excess flow to its neighbors. 

Given $x_{u}$ as excess flow of node $u$, $push(u,v)$ operation will send a flow of amount $\Delta=min(x_{u},f_{uv})$ from node $u$ to node $v$ and then will decrease excess flow of $u$ by $\Delta$ (i.e., $x_{u}=x_{u}-\Delta$) and will increase excess flow of $v$ by $\Delta$ (i.e., $x_{v}=x_{v}+\Delta$). After $push(u,v)$ operation, node $v$ will be put in a queue of active nodes to be considered by the push-relabel algorithm in the next iterations and residual graph would be updated. Figure \ref{fig:push_ex:c} shows the result of $push(u,v)$ and $push(u,k)$ on the graph shown in Figure \ref{fig:push_ex:b}. In $push(u,v)$, for instance, since $u$ and all of its neighbors have the same label value, in order to perform push operation, first we need to perform relabel operation on node $u$ to increase the label of $u$ by one unit more than the minimum label of its neighbors to guaranty that there is at least one neighbor with lower label for performing push operation. After that, node $u$ can send flow to its neighbors. 

Given $x_{u}=15$, $f_{uv}=8$, and $f_{uk}=4$ in Figure \ref{fig:push_ex:b}, after performing relabel operation, we can only send the flow of amount 8 from $u$ to $v$ and the flow of amount 4 from $u$ to $k$. After these operations, residual graph (backward flow from $v$ and $k$ to $u$) will be updated. 

The push-relabel algorithm starts with a "preflow" operation to initialize the variables and then it iteratively performs push or relabel operations until no active node exists for performing operations. Assuming $\mathcal{L}_v$ as the label of node $v$, in preflow step, we initialize all nodes as follow: $\mathcal{L}_{s_1}=|I|+|U|+2$, $\mathcal{L}_{i \in I}=2$, $\mathcal{L}_{u \in U}=1$, and $\mathcal{L}_{s_2}=0$. This way, we will be able to send the flow from $s_1$ to $s_2$ as the left nodes have higher label than the right nodes. Also, we will push the flow of amount $w_{s_{1}i}$ (where $i \in I$) from $s_1$ to all the left nodes. 
After preflow, all of the left nodes $i \in I$ will be in the queue, $\mathcal{Q}$, as active nodes because all those nodes now have positive excess flow. The main part of the algorithm will now start by dequeuing an active node $v$ from $\mathcal{Q}$ and performing either push or relabel operations on $v$ as explained above. This process will continue until $\mathcal{Q}$ is empty. At the end, each node will have specific label value and the sum of all the coming flows to node $s_2$ would be the maximum flow of graph $G$. For more details see \cite{goldberg1988}\footnote{You can interactively run Push-relabel algorithm on your graph here: http://www.adrian-haarbach.de/idp-graph-algorithms/implementation/maxflow-push-relabel/index\_en.html.}. 

An important question is: \textit{how does the Push-relabel algorithm can find high-quality (more relevant) nodes (items and their suppliers) with low degree (visibility)?} 
We answer this question by referring to the example in Figure \ref{fig:push_ex:c}. In this figure, assume that $u$ has a backward edge to $s_1$. Since $u$ has excess flow greater than zero, it should send it to its neighbors. However, as you can see in the figure, $u$ does not have any forward edge to $v$ or $k$ nodes. Therefore, it has to send its excess flow back to $s_1$ as $s_1$ is the only reachable neighbor for $u$. Since $s_1$ has the highest label in our setting, in order for $u$ to push all its excess flow back to $s_1$, it should go through a relabel operation so that its label becomes larger than that of $s_1$. Therefore, the label of $u$ will be set to $\mathcal{L}_{s_1}+1$ for an admissible push. 

The reason that $u$ receives high label value is the fact that it initially receives high flow from $s_1$ (it is important how to assign weight to edges between $s_1$ and left nodes), but it does not have enough capacity (the sum of weights between $u$ and its neighbors is smaller than its excess flow. i.e. 8+4<15) to send all that flow to them. 

In FairMatch, in step 2 (i.e. section \ref{weight_comp}), the same weight is assigned to all edges connected to $s_1$ and $s_2$. This means that the capacity of edges from $s_1$ to all item nodes would be the same and also the capacity of edges from all user nodes to $s_2$ would be the same. However, the weights assigned to edges between item and user nodes depend on the quality and visibility of the recommended items to users in the recommendation lists and play an important role in finding the desired output in FairMatch algorithm. Assume that the weights for edges between $s_1$ and item nodes are $w_{s_1}$ and the weights for edges between user nodes and $s_2$ are $w_{s_2}$.

In preflow step, $s_1$ sends flow of amount $w_{s_1}$ to each item node in $I$ and this flow would be recorded in each item nodes as their excess flow. When Push-relabel starts after preflow, the algorithm tries as much as possible to send the excess flow in item nodes to user nodes and then finally to $s_2$. However, the possibility of achieving this objective depends on the capacity of edges between item and user nodes. Items connected to edges with low capacity will not be able to send all their excess flow to their neighbors (user nodes) and will be returned as candidate items in step 3 of FairMatch algorithm. 

There are two possible reasons for some items to not be able to send all their excess flow to their neighbors: 1) they have few neighbors (user nodes) which signifies that those items are recommended to few users and consequently they have low visibility in recommendation lists, 2) they are relevant to the users' preferences meaning that their rank in the recommendation list (sorted based on the predicted score by a base recommender) for users is low and consequently make those items more relevant to users. Hence, these two reasons--low visibility and high relevance--cause some items to not have sufficient capacity to send their excess flow to their neighbors (user nodes) and have to send it back to $s_1$ similar to what we illustrated above. As a result, sending back the excess flow to $s_1$ means first running relabel operation as $s_1$ has higher label value than item nodes and then push the excess flow to $s_1$. Performing relabel operation will assign the highest label value to those items which makes them to be distinguishable from other nodes after push-relabel algorithm terminated. Therefore, in step 3 (i.e. section \ref{fair_max}), left nodes without sufficient capacity on their edges will be returned as part of the outputs from push-relabel algorithm and are considered for constructing the final recommendation list in step 4 (i.e. section \ref{step4}). FairMatch aims at promoting those high relevance items (or suppliers) with low visibility.

\subsection{Recommendation List Construction}\label{step4}

In this step, the goal is to construct a recommendation list of size $n$ by the $<user,item>$ pairs identified in previous step. Given a recommendation list of size $n$ for user $u$, $L_{u}$, sorted based on the scores generated by a base recommendation algorithm, candidate items identified by FairMatch connected to $u$ as $\mathcal{I}_{C}$, and visibility of each item $i$ in recommendation lists of size $n$ as $\mathcal{V}_{i}$, we use the following process for generating recommendation list for $u$. First, we sort recommended items in $L_{u}$ and $\mathcal{I}_{C}$ based on their $\mathcal{V}_{i}$ in ascending order. Then, we remove $min(\beta \times n,|\mathcal{I}_{C}|)$ from the bottom of sorted $L_{u}$ and add $min(\beta \times n,|\mathcal{I}_{C}|)$ items from $\mathcal{I}_{C}$ to the end of $L_{u}$. $\beta$ is a hyperparameter in $0<\beta \leq 1$ that specifies the fraction of items in the original recommendation lists that we want to replace with the identified items in previous step.

This process will ensure that extracted items in the previous step will replace the frequently recommended items meaning that it decreases the visibility of the frequently recommended items/suppliers and increases the visibility of rarely recommended items/suppliers to generate a fairer distribution on recommended items/suppliers.

\section{Experimental Methodology}

We performed a comprehensive evaluation of the effectiveness of FairMatch algorithm in improving the exposure fairness of recommender systems. Our evaluation on three standard recommendation algorithms and comparison to various bias mitigation methods as baselines on two publicly available datasets shows that FairMatch algorithm significantly improves the exposure fairness of the recommendations with a relatively small loss in the accuracy of recommendations. 

\subsection{Data}

Experiments are performed on two publicly available datasets: Last.fm\footnote{http://www.cp.jku.at/datasets/LFM-1b/} \cite{schedl2016lfm} and MovieLens \cite{harper2015}. Last.fm dataset contains user interactions with songs (and the corresponding albums). We used the same methodology in \cite{kowald2020unfairness} to turn the interaction data into rating data using the frequency of the interactions with each item (more interactions with an item will result in higher rating). In addition, we used albums as the items to reduce the size and sparsity of the item dimension, therefore the recommendation task is to recommend albums to users. We considered the artists associated with each album as the supplier of that album. After pre-processing the data\footnote{We removed users with less than 50 ratings and items less than 200 ratings to create a denser dataset and then, we randomly sampled 2,000 users from the data.}, there are 2,000 users who provided 218,985 ratings on 6,817 albums. Also, there are 2,856 suppliers (i.e. artists) in this dataset. 

\begin{figure*}[btp]
    \centering
    \begin{subfigure}[b]{0.45\textwidth}
        \includegraphics[width=\textwidth]{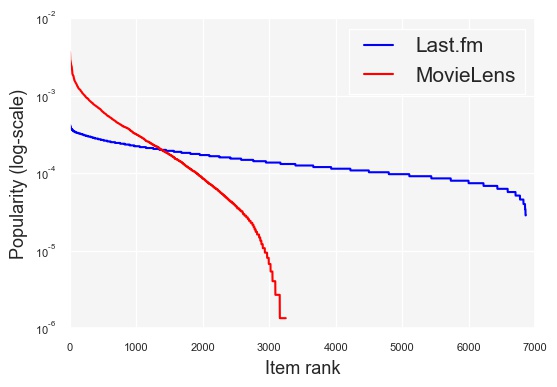}
        \caption{Last.fm} \label{}
    \end{subfigure}
\caption{Item popularity in Last.fm and MovieLens datasets.} \label{train_dist}
\end{figure*}

The MovieLens dataset is a movie rating data and was collected by the GroupLens\footnote{https://grouplens.org/datasets/movielens/} research group. We considered the movie-maker associated with each movie as the supplier of that movie. Since this dataset does not originally contain information about the movie-makers, we used the API provided by OMDB (not to be confused with IMDB) website\footnote{http://www.omdbapi.com/} to extract the information about movie-makers associated with different movies. Overall, there are 6,040 users who provided 928,739 ratings on 3,079 movies and there are 1,699 suppliers (i.e. movie-makers) in this dataset.

These datasets are from different domains, have different levels of sparsity, and are different in terms of popularity distribution of different items. Figure \ref{train_dist} shows the distribution of item popularity in both datasets. For readability, we used log-scale for the y-axis. We can see that, MovieLens shows a much more extreme long-tail shape than Last.fm indicating there are more inequality in terms of the number of times each item is rated. In Last.fm, generally the distribution seems to be fairer for different items which means there is less bias in data and therefore less bias will be in the recommendations. That, however, also means that the reranking algorithms will have a harder time improving the visibility of different items (as we will see in Table \ref{tlb_lf_50} and Figures \ref{k_a_lf} and \ref{k_sa_lf}) since the base algorithm already does a relatively good job in doing so. 

\begin{figure*}[btp]
    \centering
    \begin{subfigure}[b]{0.49\textwidth}
        \includegraphics[width=\textwidth]{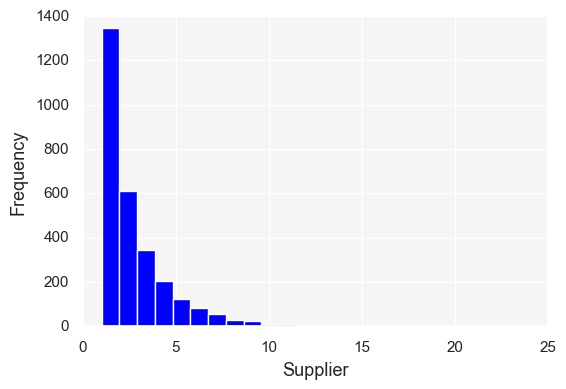}
        \caption{Last.fm} \label{lf-hist}
    \end{subfigure}
    \begin{subfigure}[b]{0.49\textwidth}
        \includegraphics[width=\textwidth]{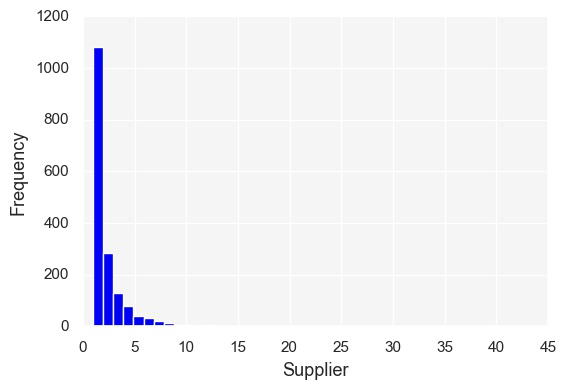}
        \caption{MovieLens} \label{ml-hist}
    \end{subfigure}%
\caption{Histogram of suppliers inventory (number of items each supplier owns).} \label{hist}
\end{figure*}

Also it is worth noting that different suppliers do not own the same number of items as we can see in Figure \ref{hist} where the majority of suppliers have only one item. Because of this, we will see that both versions of our FairMatch algorithm perform relatively similar in some cases since improving item visibility for those items that belong to suppliers with only one item is indeed equivalent to improving the visibility of the corresponding supplier. 

\subsection{Setup}

We used 80\% of each dataset as our training set and the other 20\% for the test. The training set was used for building a recommendation model and generating recommendation lists, and the test set was used for evaluating the performance of generated recommendations. We generated recommendation lists of size $t=50$ for each user using each recommendation algorithm. We then extract the final recommendation lists of size $n=10$ using each reranking method by processing the recommendation lists of size 50. We also performed experiments with $t=100$ and $n \in \{5,15,20\}$, but in this paper, we only report the results for $t=50$ and $n=10$ as we observed the same patterns in all experiments. We used \textit{librec-auto} and LibRec 2.0 for running the experiments \cite{mansoury2019algorithm,Guo2015}. 

The initial longer recommendation lists of size $t=50$ are generated by three well-known recommendation algorithms: Bayesian Personalized Ranking (\algname{BPR}) \cite{rendle2009bpr}, Neural Collaborative Filtering (\algname{NCF}) \cite{he2017neural}, and User-based Collaborative Filtering (\algname{UserKNN}) \cite{Resnick:1994a}. We chose these algorithms to cover different approaches in recommender systems: matrix factorization, neural networks, and neighborhood models. We performed gridsearch\footnote{For \algname{BPR}, we set all regularizers $\in \{0.0001,0.001,0.01\}$, $iterations \in \{30,50,100\}$, $learning rate \in \{0.0001,0.001,0.005,0.01\}$, and $factors \in \{50,100,150,200\}$. For \algname{NCF}, we set $epochs \in \{10,20\}$, $factors \in \{8, 15, 30\}$, and $learning rate \in \{0.001, 0.0001\}$. For \algname{UserKNN}, we set $neighbors \in \{10, 30, 50, 100, 200, 300\}$.} on hyperprameters for each algorithm and selected the results with the highest precision value for our next analysis. We used McNemar's test to evaluate the significance of results and based on McNemar's test, the results shown with underline in Tables \ref{tlb_lf_50} and \ref{tbl_ml_50} are statistically significant with a p-value of less than 0.05.

\subsection{Baselines}\label{baselines}

To show the effectiveness of the FairMatch algorithm in improving the exposure fairness of recommendations, we compare its performance with four state-of-the-art algorithms and also two simple baselines. 

\begin{enumerate}
  \item \textbf{FA*IR.} This is the method introduced in \cite{zehlike2017fa} and was mentioned in our related work section. The method was originally used for improving the representation of protected group in ranked recommendation lists. However, we use this method for improving the visibility of long-tail items in recommendation lists. We define protected and unprotected groups as long-tail and short-head items, respectively. For separating short-head from long-tail items, we consider those top items which cumulatively take up 20\% of the ratings according the Pareto principle \cite{sanders1987pareto} as the short-head and the rest as long-tail items. Also, we set the other two hyperparameters, proportion of protected candidates in the top $n$ items\footnote{Based on suggestion from the released code, the range should be in $[0.02,0.98]$} and significance level\footnote{Based on suggestion from the released code, the range should be in $[0.01,0.15]$}, to $\{0.2,0.6,0.8\}$ and $\{0.05,0.1\}$, respectively. 
  \item \textbf{xQuAD}. This is the method introduced in \cite{abdollahpouri2019managing} and was mentioned in our related work section. We specifically included xQuAD method since it attempts to promote less popular items (most likely items with low visibility in recommendation lists) by balancing the ratio of popular and less popular items in recommendation lists. This method involves a hyperparameter to control the trade-off between relevance and long-tail promotion, and we experimented with different values for this hyperparameter in $\{0.2,0.4,0.6,0.8,1\}$. Also, the separation of short-head and long-tail items is done according to Pareto principle as described above.
  \item \textbf{Discrepancy Minimization (DM)}. This is the method introduced in \cite{antikacioglu2017} and was explained in our related work section. For hyperparameter tuning, we followed the experimental settings suggested by the original paper for our experiments. We set the target degree distribution to $\{1,5,10\}$ and relative weight of the relevance term to $\{0.01,0.5,1\}$.  
  \item \textbf{ProbPolicy}. This is the method introduced in \cite{mehrotra2018towards} and was mentioned in our related work section. We included this method as it was designed for improving supplier fairness and visibility in recommendation lists. This method involves a hyperparameter for controlling the trade-off between the relevance of recommended items to users and supplier fairness. We set the value for this hyperparameter to $\{0.2,0.4,0.6,0.8,1\}$.
\end{enumerate}

We also used two simple methods to show the extreme case in bias mitigation for comparison purposes.
\begin{enumerate}

  \item \textbf{Reverse.} Given a recommendation list of size $t$ for each user generated by base recommendation algorithm, in this method, instead of picking the $n$ items from the top (most relevant items), we pick them from the bottom of the list (least relevant items). In this approach, we expect to see an increase in aggregate diversity as we are giving higher priority to the items with lower relevance to be picked first. However, the accuracy of the recommendations will decrease as we give higher priority to less relevant items.
  \item \textbf{Random.} Given a recommendation list of size $t$ for each user generated by base recommendation algorithm, we randomly choose $n$ items from that list and create a final recommendation list for that user. Note that this is different from randomly choosing items from all catalog to recommend to users. The reason we randomly choose the items from the original recommended list of items (size $t$) is to compare other post-processing and re-ranking techniques with a simple random re-ranking. 
\end{enumerate}

\textit{Random} and \textit{Reverse} are mainly included to demonstrate the extreme version of a re-ranking algorithm where our sole focus is on improving aggregate diversity and exposure and we ignore the relevance of the recommended items as can be seen by the low precision for these two algorithms.

One of the hyperparameters involved in our FairMatch algorithm is $\lambda$ which controls the balance between the utility of users and suppliers (or items). For our experiments we try $\lambda \in \{0,0.25,0.5,0.75,1\}$. A higher value for $\lambda$ indicates more focus on maintaining the accuracy of the recommendations, while a lower value for $\lambda$ indicates more focus on improving exposure fairness of recommendations. Another hyperparameter in FairMatch algorithm is $\beta$ which determines the fraction of items in original recommendation lists that we want to replace with the identified items in the third step of FairMatch algorithm. For our experiments, we set $\beta=0.6$ and $\beta=1$ for Last.fm and MovieLends datasets, respectively. 

\subsection{Evaluation metrics}\label{metrics}

For evaluation, we use the following metrics to measure different aspects of the effectiveness of each method:

\begin{enumerate}
    \item \textbf{Precision ($P$)}: The fraction of the recommended items shown to the users that are part of the users' profile in the test set.
    \item \textbf{Item Visibility Shift ($IVS$)}: The percentage of increase or decrease in the visibility of item groups in final recommendation lists generated by a reranking algorithm compared to their visibility in recommendation lists generated by a base recommender. Given long recommendation lists of size $t$, $L'$, generated by a base recommender and the visibility of each item $i$ computed as the fraction of times that it appears in the recommendation lists of different users 
    , we create 10 groups of items based on their visibility in $L'$. To do so, first, we sort the recommended items based on their visibility in $L'$ in descending order, and then we group the recommended items into 10 equal-sized bins where the first group represents the items with the highest visibility and 10th group represents the items with the lowest visibility in $L'$. 
    Item Visibility ($IV$) of each item $i$ in final recommendation lists can be calculated as:
    
    \begin{equation}\label{v_i}
        IV(i)=\frac{\sum_{j \in L}\mathds{1}(j=i)}{|L|}
    \end{equation}
    
    where $\mathds{1}(.)$ is the indicator function returning zero when its argument is False and 1 otherwise. Item Group Visibility ($IGV$) for each item group $\tau$ can be calculated as:
    
    \begin{equation}
        IGV({\tau})=\frac{\sum_{i \in \tau}{IV(i)}}{|\tau|}
    \end{equation}
    
    Therefore, Item Visibility Shift ($IVS$) of group $\tau$ can be calculated as:
    
    \begin{equation}\label{IVS_g}
        IVS(\tau)= \frac{IGV(\tau)^{Reranker}-IGV(\tau)^{Base}}{IGV(\tau)^{Base}}    
    \end{equation}
    
    where $IGV(\tau)^{Reranker}$ and $IGV(\tau)^{Base}$ are the visibility of item group $\tau$ in recommendation lists of size $n$ generated by reranking algorithm and the base algorithm, respectively.
    
    \item \textbf{Supplier Visibility Shift ($SVS$)}: The percentage of increase or decrease in the visibility of supplier groups in final recommendation lists generated by a reranking algorithm compared to their visibility in recommendation lists generated by a base recommender. $SVS$ can be calculated similar to $IVS$, but instead of calculating the percentage change over item groups, we calculate it over supplier groups in $SVS$. Thus, given long recommendation lists $L'$ generated by a base recommender and the visibility of each supplier $s$ computed as the fraction of times the items belonging to that supplier appear in the recommendation lists of different users, analogous to $IVS$, we create 10 groups of suppliers based on their visibility in $L'$. 
    Supplier Visibility ($SV$) of each supplier $s$ in final recommendation lists $L$ can be calculated as:
    
    \begin{equation}\label{SV(s)}
        SV(s)=\sum_{s \in g}{\sum_{i \in A(s)}{IV(i)}}
    \end{equation}
    
    where $A(s)$ returns the items belonging to supplier $s$. Supplier Group Visibility ($SGV$) for each supplier group $g$ can be calculated as:
    
    \begin{equation}
    SGV(g)=\frac{SV(s)}{|g|}
    \end{equation}
    
    Therefore, Supplier Visibility Shift ($SVS$) of group $g$ can be calculated as:
    
    \begin{equation}\label{IVS_g}
        SVS(g)= \frac{SGV(g)^{Reranker}-SGV(g)^{Base}}{SGV(g)^{Base}}    
    \end{equation}
    
    where $SGV(g)^{Reranker}$ and $SGV(g)^{Base}$ are the visibility of item group $g$ in recommendation lists of size $n$ generated by reranking algorithm and the base algorithm, respectively.

    \item \textbf{Item Aggregate Diversity ($\alpha\mbox{-}IA$)}: We propose $\alpha\mbox{-}IA$ as the fraction of items which appear at least $\alpha$ times in the recommendation lists and can be calculated as:
    
    \begin{equation}
        \alpha\mbox{-}IA=\frac{\sum_{i \in I}\mathds{1}{(\sum_{j \in L}\mathds{1}(j=i) \geq \alpha)}}{|I|} ,\quad (\alpha \in \mathbb{N})
    \end{equation}

    This metric is a generalization of standard aggregate diversity as it is used in \cite{vargas2011rank,adomavicius2011maximizing} where $\alpha=1$.
    
    \item \textbf{Long-tail Coverage ($LT$):} The fraction of the long-tail items covered in the recommendation lists. To determine the long-tail items, we separated the top items which cumulatively take up 20\% of the ratings in train data as short-head and the rest of the items are considered as long-tail items. Given these long-tail items, we calculated $LT$ as the fraction of these items appeared in recommendation lists.
    
    \item \textbf{Supplier Aggregate Diversity ($\alpha\mbox{-}SA$)}: We propose $\alpha\mbox{-}SA$ as the fraction of suppliers which appear at least $\alpha$ times in the recommendation lists and can be calculated as:
    
    \begin{equation}
        \alpha\mbox{-}SA=\frac{\sum_{s \in S}\mathds{1}{(\sum_{i \in A(s)}\sum_{j \in L}\mathds{1}(j=i) \geq \alpha)}}{|S|} ,\quad (\alpha \in \mathbb{N})
    \end{equation}
    
    where $A(s)$ returns all the items belonging to supplier $s$ and $S$ is the set of all suppliers. 
   
    \item \textbf{Item Gini Index ($IG$)}: The measure of fair distribution of recommended items. It takes into account how uniformly items appear in recommendation lists. Uniform distribution will have Gini index equal to zero which is the ideal case (lower Gini index is better). $IG$ is calculated as follows over all the recommended items across all users:
    
    \begin{equation}\label{IG(L)}
        IG=\frac{1}{|I|-1} \sum_{k=1}^{|I|} (2k-|I|-1)IV(i_{k}) 
    \end{equation}
    
    where $IV(i_k)$ is the visibility of the $k$-th least recommended item being drawn from $L$ and is calculated using Equation \ref{v_i}.
    
    \item \textbf{Supplier Gini Index ($SG$)}: The measure of fair distribution of suppliers in recommendation lists. This metric can be calculated similar to $IG$, but instead of considering the distribution of recommended items, we consider the distribution of recommended suppliers and it can be calculated as:
    
    \begin{equation}\label{IG(L)}
        SG=\frac{1}{|S|-1} \sum_{k=1}^{|S|} (2k-|S|-1)SV(s_{k}) 
    \end{equation}
    
    where $SV(s_{k})$ is the visibility of the $k$-th least recommended supplier being drawn from $L$ and is calculated using Equation \ref{SV(s)}. 
    
    \item \textbf{Item Entropy ($SE$)}: Given the distribution of recommended items, entropy measures the uniformity of that distribution. Uniform distribution has the highest entropy or information gain, thus higher entropy is more desired when the goal is increasing diversity. 
    
    \begin{equation}
        IE=- \sum_{i \in I}^{} {IV(i) }\log IV(i)
    \end{equation}
    
    \item \textbf{Supplier Entropy ($SE$)}: The measure of uniformity of the distribution of suppliers in the recommendation lists. Similar to Gini where we had both $IG$ and $SG$, we can also measure the entropy for suppliers as follows:
    
    \begin{equation}
        SE=- \sum_{s \in S} SV(s) \log SV(s)
    \end{equation}
\end{enumerate}

\begin{figure*}[btp]
    \centering
    \begin{subfigure}[b]{0.99\textwidth}
        \includegraphics[width=\textwidth]{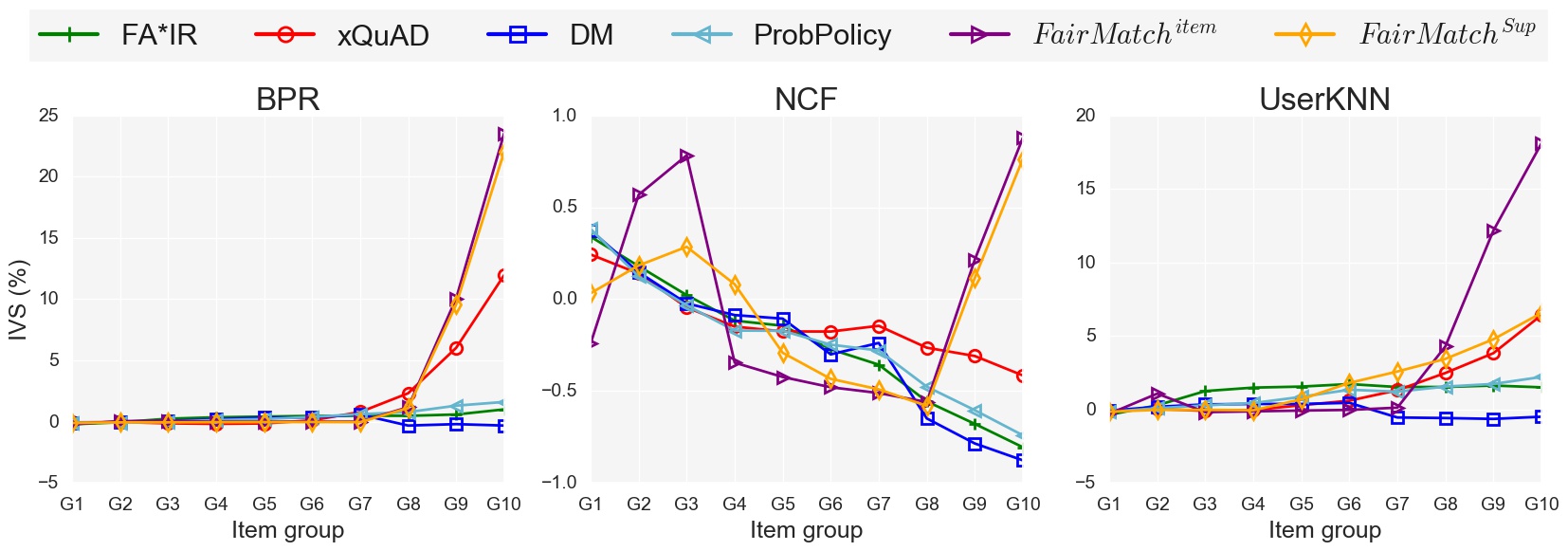}
        \caption{Last.fm} \label{vis_lf_50_item}
    \end{subfigure}
    \begin{subfigure}[b]{0.99\textwidth}
        \includegraphics[width=\textwidth]{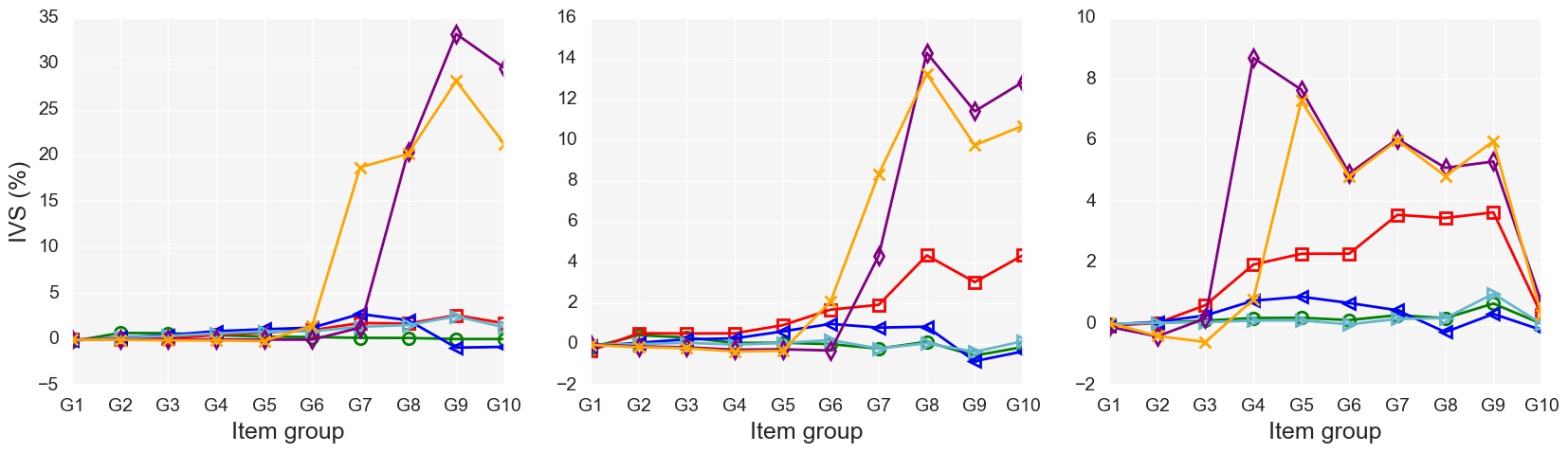}
        \caption{MovieLens} \label{vis_ml_50_item}
    \end{subfigure}%
\caption{Percentage increase/decrease ($IVS$) in visibility of item groups for different reranking algorithms.} \label{vis_50_item}
\end{figure*}

\section{Results}

In this section, we analyze the performance of our \textit{FairMatch} algorithm in comparison with some of the state-of-the-art reranking algorithms we described in Section \ref{baselines} using three different standard recommendation algorithms as the base for the reranking algorithms on two datasets. Extensive experiments are performed using each reranking algorithm with multiple hyperparameter values. For the purpose of fair comparison, from each of those reranking algorithms (FA*IR, xQuAD, DM, ProbPolicy, and the both variations of our FairMatch algorithm) the configuration
which yields, more of less, the same precision loss is reported. These results enable us to better compare the performance of each technique on improving exposure fairness and other non-accuracy metrics while maintaining the same level of accuracy. The precision of each re-ranking algorithm on both datasets is reported in Tables \ref{tlb_lf_50} and \ref{tbl_ml_50}. 

\begin{figure*}[btp]
    \centering
    \begin{subfigure}[b]{0.99\textwidth}
        \includegraphics[width=\textwidth]{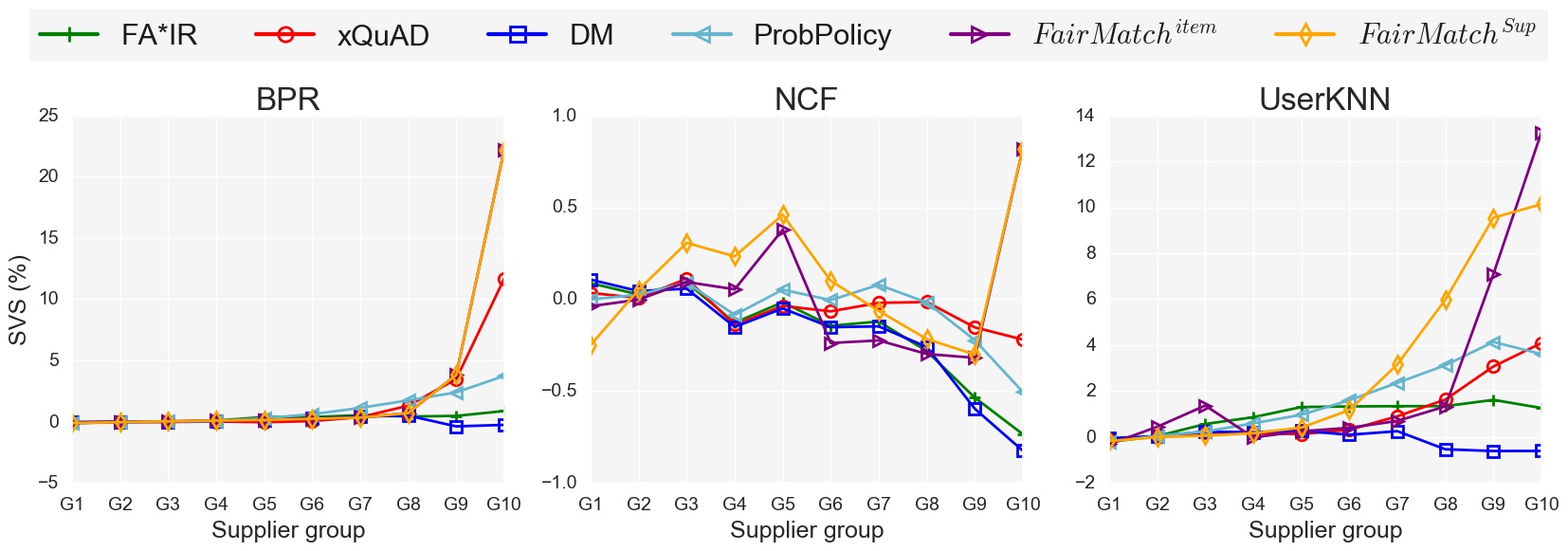}
        \caption{Last.fm} \label{vis_lf_50_sup}
    \end{subfigure}
    \begin{subfigure}[b]{0.99\textwidth}
        \includegraphics[width=\textwidth]{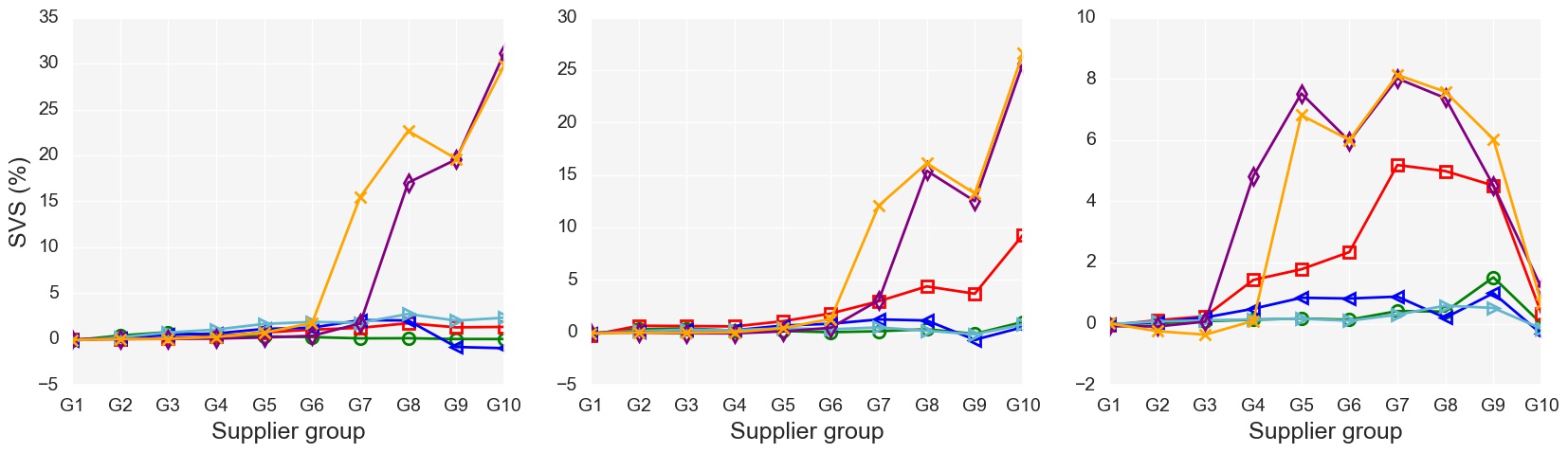}
        \caption{MovieLens} \label{vis_ml_50_sup}
    \end{subfigure}%
\caption{Percentage increase/decrease ($SVS$) in visibility of supplier groups for different reranking algorithms.} \label{vis_50_sup}
\end{figure*}

\subsection{Visibility Analysis}

Since our FairMatch algorithm aims at improving the visibility of different items in the recommendations we start our analysis with comparing different algorithms in terms of the visibility change ($IVS$) of the recommended items. Figure \ref{vis_50_item} shows the percentage change in the visibility of the recommended item groups in recommendation lists generated by each re-ranking algorithm compared to their visibility in the recommendation lists generated by three base recommenders. In these plots, x-axis is the recommended item groups (created as explained in section \ref{metrics}) and y-axis is $IVS$ metric. Item groups are sorted from the highest visibility (i.e., $G_1$) to the lowest visibility (i.e., $G_{10}$). It can be seen that both versions of FairMatch algorithm on both datasets have significantly increased the visibility of item groups with lower visibility while slightly taking away from the visibility of items with originally extreme visibility. $FairMatch^{item}$ performs slightly better than $FairMatch^{Sup}$ especially for item groups for very low visibility ($G_9$ and $G_{10}$) as it was expected since $FairMatch^{item}$ directly optimizes for improving the exposure of the low visibility items. 

Looking at \algname{NCF} on MovieLens, it seems different reranking algorithms do not have a predictable behavior in terms of improving visibility of different item groups and, in some cases, even decreasing the visibility of item groups with already low visibility. However, a closer look at the scale of y-axis reveals that these changes are very small and not significant. The reason is, on this dataset, \algname{NCF} has already done a good job in terms of fair item visibility and not much can be done via a reranking method. Among other reranking methods, xQuAD seems to also perform relatively well but still is outperformed by FairMatch. One interesting observation in this figure is that, using \algname{UserKNN} on MovieLens, we can see that both FairMatch algorithms have significantly improved the visibility of item groups with medium visibility even more than the ones with lower visibility. Although these are items with medium visibility using our grouping strategy, they still get significantly less visibility compared to $G_1$ and $G_2$ in the base algorithm as we saw in Figure \ref{dist_ml_50_item}. Therefore, we can still consider these item groups as items with relatively low visibility and FairMatch has increased their visibility.   

Figure \ref{vis_50_sup} is similar to \ref{vis_50_item} but here we show the percentage change in the visibility of the supplier groups in recommendation lists generated by each reranking algorithm compared to their visibility in the recommendation lists generated by three base recommenders. The first thing that can be observed from this figure is that, on both datasets, FairMatch algorithms outperform the other reranking methods especially for groups with lower visibility. \algname{NCF} on Last.fm has the same problem as we observed in Figure \ref{vis_50_item} where the changes in y-axis are not significant and all algorithms more or less perform equally. $FairMatch^{item}$ and $FairMatch^{Sup}$ are performing equally well for groups with extremely low visibility although $FairMatch^{Sup}$ tend to also improve the visibility of some other item groups such as $G_7$, $G_8$, and $G_9$ on MovieLens using \algname{BPR} and on Last.fm using \algname{UserKNN}. Overall, $FairMatch^{Sup}$ has done a better job in terms of supplier visibility fairness and that was also expected since we incorporated supplier visibility directly into our objective function.  

In addition to measuring the improvement in visibility of different items, we also conducted an extensive analysis on other existing metrics in the literature to have a better picture of how each of these reranking methods help reducing the over-concentration of the recommendations around few highly visible items. Table \ref{tlb_lf_50} and \ref{tbl_ml_50} show the results for different reranking algorithms on Last.fm and MovieLens datasets, respectively. We compare these algorithms in terms of item and supplier aggregate diversity and also fair distribution of recommended items and suppliers.

\captionsetup[table]{skip=4pt}
\begin{table*}[t]
\centering
\setlength{\tabcolsep}{3pt}
\captionof{table}{
Comparison of different reranking algorithms on \textbf{Last.fm} dataset for long recommendation lists of size 50 ($t=50$) and final recommendation lists of size 10 ($n=10$). The bolded entries show the best values and the underlined entries show the statistically significant change from the second-best baseline with $p<0.05$ (comparison between FairMatch algorithms and other baselines ignoring Random and Reverse).}
 \label{tlb_lf_50}
\begin{tabular}{llrrrrrrrrrrrrr}
\toprule

 algorithms & baselines & Precision & $1\mbox{-}IA$ & $5\mbox{-}IA$ & $LT$ & $1\mbox{-}SA$ & $5\mbox{-}SA$ & $IG$ & $IE$ & $SG$ & $SE$ \\
 \bottomrule
 
 \multirow{9}{*}{\algname{BPR}}   
 & Base & 0.097 & 0.555 & 0.218 & 0.53 & 0.668 & 0.374 & 0.693 & 7.83 & 0.686 & 7 \\
 & Random & 0.062 & 0.695 & 0.237 & 0.678 & 0.781 & 0.424 & 0.568 & 8.16 & 0.607 & 7.23 \\
 & Reverse & 0.041 & 0.768 & 0.243 & 0.755 & 0.847 & 0.455 & 0.492 & 8.31 & 0.564 & 7.33 \\
 & FA*IR & 0.096 & 0.613 & \underline{\textbf{0.242}} & 0.591 & 0.715 & \textbf{0.421} & 0.627 & 8.01 & 0.642 & 7.13 \\ 
 & xQuAD & 0.094 & 0.677 & 0.188 & 0.659 & 0.787 & 0.373 & 0.646 & 7.95 & 0.653 & 7.1 \\
 & DM & 0.096 & 0.644 & 0.221 & 0.625 & 0.736 & 0.399 & 0.627 & 8.01 & 0.649 & 7.11 \\
 & ProbPolicy & 0.092 & 0.607 & 0.22 & 0.586 & 0.784  & 0.419 & 0.659 & 7.93 & 0.618 & 7.19 \\
 & $FairMatch^{item}$ & 0.092 & \textbf{0.686} & 0.223 & \textbf{0.669} & \textbf{0.791} & 0.404 & \underline{\textbf{0.602}} & \textbf{8.08} & 0.623 & 7.19 \\
 & $FairMatch^{Sup}$ & 0.095 & 0.675 & 0.21 & 0.657 & \textbf{0.791} & 0.404 & 0.623 & 8.03 & \textbf{0.617} & \textbf{7.22} \\
 \hline
 
 \multirow{9}{*}{\algname{NCF}}   
 & Base & 0.08 & 0.638 & 0.211 & 0.62 & 0.754 & 0.4 & 0.666 & 7.95 & 0.661 & 7.12 \\
 & Random & 0.056 & 0.74 & 0.221 & 0.726 & 0.824 & 0.441 & 0.552 & 8.22 & 0.592 & 7.28 \\
 & Reverse & 0.044 & 0.791 & 0.227 & 0.779 & 0.857 & 0.463 & 0.492 & 8.34 & 0.561 & 7.35 \\
 & FA*IR & 0.079 & 0.653 & 0.21 & 0.639 & 0.768 & 0.41 & 0.639 & 8.01 & 0.639 & 7.16 \\ 
 & xQuAD & 0.075 & 0.694 & 0.212 & 0.683 & 0.804 & 0.424 & 0.61 & 8.08 & 0.616 & 7.22 \\ 
 & DM & 0.079 & 0.723 & 0.205 & 0.707 & 0.808 & 0.412 & 0.594 & 8.11 & 0.624 & 7.2 \\ 
 & ProbPolicy & 0.072 & 0.659 & 0.207 & 0.643 & 0.809 & 0.43 &  0.647 & 7.98 & 0.611 & 7.22 \\ 
 & $FairMatch^{item}$ & 0.064 & \textbf{0.729} & 0.224 & \textbf{0.716} & 0.821 & 0.42 & 0.589 & \textbf{8.15} & 0.608 & 7.25 \\ 
 & $FairMatch^{Sup}$ & 0.071 & 0.711 & \underline{\textbf{0.229}} & 0.699 & \underline{\textbf{0.828}} & \underline{\textbf{0.485}} & \textbf{0.588} & \textbf{8.15} & \underline{\textbf{0.535}} & \underline{\textbf{7.41}} \\ 
 \hline
 
 \multirow{9}{*}{\algname{UserKNN}}   
 & Base & 0.08 & 0.461 & 0.127 & 0.431 & 0.588 & 0.257 & 0.833 & 7.07 & 0.813 & 6.39 \\
 & Random & 0.044 & 0.635 & 0.164 & 0.615 & 0.738 & 0.341 & 0.689 & 7.77 & 0.702 & 6.91 \\
 & Reverse & 0.027 & 0.712 & 0.204 & 0.696 & 0.797 & 0.394 & 0.591 & 8.09 & 0.634 & 7.14 \\
 & FA*IR & 0.074 & \textbf{0.629} & \underline{\textbf{0.172}} & \textbf{0.609} & 0.73 & \textit{0.351} & \underline{\textbf{0.687}} & \textbf{7.73} & \textit{0.706} & \textit{6.86} \\  
 & xQuAD & 0.078 & 0.577 & 0.117 & 0.554 & 0.701 & 0.269 & 0.781 & 7.31 & 0.771 & 6.58 \\ 
 & DM & 0.077 & 0.537 & 0.134 & 0.512 & 0.638 & 0.283 & 0.782 & 7.33 & 0.781 & 6.55 \\
 & ProbPolicy & 0.067 & 0.559 & 0.14 & 0.535 & 0.785 & \underline{\textbf{0.332}} & 0.765 & 7.4 & 0.7 & 6.84 \\ 
 & $FairMatch^{item}$ & 0.062 & 0.626 & 0.102 & 0.606 & 0.751 & 0.266 & 0.745 & 7.51 & 0.736 & 6.78 \\ 
 & $FairMatch^{Sup}$ & 0.073 & \textbf{0.629} & 0.123 & \textbf{0.609} & \underline{\textbf{0.895}} & 0.266 & 0.73 & 7.54 & \underline{\textbf{0.659}} & \textbf{6.97} \\ 
 
 \bottomrule
\end{tabular}
\end{table*}

\subsection{Item aggregate diversity}
When it comes to increasing the number of unique recommended items (aggregate diversity), we can see that all reranking algorithms have improved this metric over the base algorithms on both datasets. We have only included $1\mbox{-}IA$ (each item should be recommended at least once to be counted) and $5\mbox{-}IA$ (each item should be recommended at least 5 times to be counted). We experimented with different values of $\alpha$ from 1 to 20 and the results can be seen in Figure \ref{k_a} which we will describe afterwards. Generally speaking, all reranking methods have lost a certain degree of precision in order to improve aggregate diversity and other metrics related to fair distribution of recommended items and suppliers as can be seen from Tables \ref{tlb_lf_50} and \ref{tbl_ml_50}. The reason is that the base algorithms are mainly optimized for relevance and therefore it is more likely for the items on top of the recommended list to be relevant to the users. As a result, when we rerank the recommended lists and push some items in the bottom to go up to the top-n, we might swipe some relevant items with items that may not be as relevant.

Regarding $1\mbox{-}IA$, $FairMatch^{item}$ seems to perform relatively better than the other rerankers using all three base algorithms (\algname{BPR}, \algname{NCF} and \algname{UserKNN}) on both datasets indicating it recommends a larger number of items across all users. The same pattern can be seen for $LT$ which measures only the unique recommended items that fall into the long-tail category. This is however, not the case for $5\mbox{-}IA$ where in some cases $FairMatch^{item}$ is outperformed by other rerankers. That shows, the improvement in recommending more unique items using $FairMatch^{item}$ is not achieved by recommending them frequent enough. On MovieLens, however, $FairMatch^{item}$ performs very well on $5\mbox{-}IA$ metric. This difference in behavior across the datasets can be explained by the characteristics of the data as we saw in Figure \ref{train_dist}. Overall, $FairMatch^{Sup}$ seems to also perform relatively well on item aggregate diversity and in some cases even better than $FairMatch^{item}$ such as on $5\mbox{-}IA$ using \algname{NCF} and \algname{UserKNN} on Last.fm and \algname{BPR} and \algname{NCF} on Movielens. 

\captionsetup[table]{skip=4pt}
\begin{table*}[t]
\centering
\setlength{\tabcolsep}{3pt}
\captionof{table}{Comparison of different reranking algorithms on \textbf{MovieLens} dataset for long recommendation lists of size 50 ($t=50$) and final recommendation lists of size 10 ($n=10$). The bolded entries show the best values and the underlined entries show the statistically significant change from the second-best baseline with $p<0.05$ (comparison between FairMatch algorithms and other baselines ignoring Random and Reverse).} \label{tbl_ml_50}
\begin{tabular}{llrrrrrrrrrr}
\toprule

 algorithms & baselines & Precision & $1\mbox{-}IA$ & $5\mbox{-}IA$ & $LT$ & $1\mbox{-}SA$ & $5\mbox{-}SA$ & $IG$ & $IE$ & $SG$ & $SE$ \\
 \bottomrule
 
 \multirow{9}{*}{\algname{BPR}}   
 & Base & 0.332 & 0.392 & 0.262 & 0.351 & 0.418 & 0.276 & 0.833 & 6.01 & 0.845 & 5.4 \\
 & Random & 0.198 & 0.502 & 0.359 & 0.47 & 0.509 & 0.365 & 0.726 & 6.55 & 0.773 & 5.8 \\
 & Reverse & 0.125 & 0.547 & 0.404 & 0.518 & 0.545 & 0.394 & 0.653 & 6.81 & 0.728 & 6 \\
 & FA*IR & 0.306 & 0.402 & 0.283 & 0.362 & 0.4 & 0.296 & 0.779 & 6.33 & 0.802 & 5.5 \\ 
 & xQuAD & 0.322 & 0.461 & 0.311 & 0.426 & 0.451 & 0.324 & 0.797 & 6.19 & 0.822 & 5.34 \\
 & DM & 0.314 & 0.47 & 0.343 & 0.435 & 0.453 & 0.345 & \textbf{0.749} & \textbf{6.43} & 0.791 & \textbf{5.51} \\
 & ProbPolicy & 0.289 & 0.439 & 0.297 & 0.402 & 0.451 & 0.343 & 0.792 & 6.23 & \textit{0.782} & 5.49 \\
 & $FairMatch^{item}$ & 0.322 & 0.544 & 0.323 & 0.515 & 0.536 & 0.355 & 0.796 & 6.17 & 0.82 & 5.32 \\
 & $FairMatch^{Sup}$ & 0.311 & \underline{\textbf{0.547}} & \textbf{0.363} & \underline{\textbf{0.518}} & \underline{\textbf{0.578}} & \underline{\textbf{0.41}} & 0.768 & 6.28 & \textbf{0.779} & \textbf{5.51} \\

 \hline
 
 \multirow{9}{*}{\algname{NCF}}   
 & Base & 0.312 & 0.433 & 0.286 & 0.395 & 0.424 & 0.297 & 0.83 & 6.17 & 0.848 & 5.32 \\
 & Random & 0.198 & 0.566 & 0.396 & 0.539 & 0.571 & 0.406 & 0.728 & 6.66 & 0.776 & 5.89 \\
 & Reverse & 0.128 & 0.615 & 0.448 & 0.592 & 0.614 & 0.446 & 0.661 & 6.9 & 0.735 & 6.07 \\
 & FA*IR & 0.296 & 0.467 & 0.316 & 0.432 & 0.467 & 0.324 & 0.786 & 6.42 & 0.81 & 5.58 \\ 
 & xQuAD & 0.31 & 0.535 & 0.4 & 0.505 & 0.556 & 0.41 & 0.774 & 6.26 & \textbf{0.752} & \textbf{5.87} \\
 & DM & 0.298 & 0.53 & 0.383 & 0.499 & 0.509 & 0.386 & \textbf{0.752} & \textbf{6.54} & 0.797 & 5.6 \\
 & ProbPolicy & 0.301 & 0.469 & 0.315 & 0.434 & 0.426 & 0.344 & 0.812 & 6.24 & 0.819 & 5.45 \\
 & $FairMatch^{item}$ & 0.301 & \underline{\textbf{0.623}} & 0.375 & \underline{\textbf{0.6}} & 0.604 & 0.397 & 0.778 & 6.37 & 0.81 & 5.49 \\
 & $FairMatch^{Sup}$ & 0.292 & 0.622 & \textbf{0.407} & 0.599 & \underline{\textbf{0.62}} & \underline{\textbf{0.446}} & 0.757 & 6.45 & 0.777 & 5.66 \\
 
 \hline
 
 \multirow{9}{*}{\algname{UserKNN}}   
 & Base & 0.190 & 0.161 & 0.108 & 0.102 & 0.167 & 0.119 & 0.889 & 4.87 & 0.896 & 4.13 \\
 & Random & 0.128 & 0.219 & 0.145 & 0.164 & 0.231 & 0.157 & 0.798 & 5.51 & 0.829 & 4.79 \\
 & Reverse & 0.093 & 0.238 & 0.164 & 0.185 & 0.251 & 0.177 & 0.728 & 5.8 & 0.78 & 5.06 \\
 & FA*IR & 0.19 & 0.166 & 0.113 & 0.107 & 0.175 & 0.122 & 0.885 & 4.9 & 0.893 & 4.16 \\ 
 & xQuAD & 0.196 & 0.204 & 0.136 & 0.148 & 0.215 & 0.148 & 0.869 & 4.97 & 0.88 & 4.25 \\ 
 & DM & 0.19 & 0.183 & 0.125 & 0.125 & 0.194 & 0.133 & 0.873 & 4.98 & 0.885 & 4.22 \\ 
  & ProbPolicy & 0.19 & 0.165 & 0.11 & 0.106 & 0.175 & 0.122 & 0.886 & 4.89 & 0.891 & 4.17 \\ 
 & $FairMatch^{item}$ & 0.19 & \textbf{0.214} & \textbf{0.152} & \textbf{0.159} & \textbf{0.224} & \textbf{0.171} & \underline{\textbf{0.804}} & \textbf{5.18} & \underline{\textbf{0.831}} & \textbf{4.41} \\ 
 & $FairMatch^{Sup}$ & 0.206 & 0.207 & 0.131 & 0.152 & 0.222 & 0.158 & 0.867 & 4.77 & 0.871 & 4.15 \\ 
 
 \bottomrule
\end{tabular}
\end{table*}

\subsection{Supplier Aggregate Diversity}

Suppliers of the recommended items are also important to be fairly represented in the recommendations. First and foremost, looking at the Tables \ref{tlb_lf_50} and \ref{tbl_ml_50}, we can see that there is an overall positive connection between improving item aggregate diversity and supplier aggregate diversity indicating optimizing for either item or supplier visibility, can benefit the other side as well. However, when we directly incorporate the supplier visibility into our recommendation process as we did in $FairMatch^{Sup}$ we can see that the supplier aggregate diversity can be significantly improved. For example, we can see that $FairMatch^{Sup}$ has the best $1\mbox{-}SA$ on both datasets except for when the base algorithm is \algname{UserKNN} on MovieLens where it was outperformed by $FairMatch^{item}$. So, overall, we can say that FairMatch (either $FairMatch^{Sup}$ or $FairMatch^{item}$) has the best $1\mbox{-}SA$ on both datasets using all three base algorithms. Regarding $5\mbox{-}SA$, FairMatch algorithms tend to also perform better than other rerankers. Between the two variations of FairMatch, we can see that $FairMatch^{Sup}$ gives a better supplier aggregate diversity in most cases which is something that we expected. Similar to item aggregate diversity, we only included $1\mbox{-}SA$ and $5\mbox{-}SA$ for supplier aggregate diversity in the tables. A more comprehensive analysis of the effect of $\alpha$ on this metric is illustrated in subsection \ref{alphaanalysis} which we will describe later.

\subsection{Fair distribution of recommended items}
We also wanted to evaluate different rerankers in terms of fair distribution of recommendations across different items. We used Gini ($IG$) and Entropy ($IE$) as ways to measure how equally the recommendations are distributed across different recommended items. Even though we have not optimized directly for equal representation of different items, these two metrics show that our FairMatch algorithm has given a much fairer chance to different items to be recommended compared to the base algorithms and some of the other rerankers by having a low Gini and high Entropy. Between $FairMatch^{Sup}$ and $FairMatch^{item}$ there is no clear winner in terms of Gini and entropy for items as in some cases $FairMatch^{Sup}$ has a better Gini while in other cases $FairMatch^{item}$ performs better. Among other rerankers, DM and FA*IR seem to also perform well on these two metrics indicating they also give a fair chance to different items to be recommended. 

\subsection{Fair distribution of suppliers in recommendation lists}

In addition to standard Gini (i.e. $IG$) and Entropy (i.e. $IE$) which are generally calculated in an item level, we also measured the same metric but from the suppliers perspective and it can be seen in the tables \ref{tlb_lf_50} and \ref{tbl_ml_50} as $SG$ and $SE$ which measure the extent to which different suppliers are fairly recommended across different users. Overall, $FairMatch^{Sup}$ has the best $SG$ and $SE$ on both datasets in all situations except for \algname{NCF} and \algname{UserKNN} on Movielens. Although using \algname{UserKNN} on Movielens, $FairMatch^{item}$ has the second best $SG$ and $SE$. This shows that incorporating the supplier visibility directly into the recommendation process can positively affect the fairness of representation across different suppliers and it is indeed supporting our initial hypothesis about the importance of incorporating suppliers in the recommendation process. The Probpolicy algorithm which also incorporates the supplier fairness in its recommendation generation, has also performed better than other rerankers in terms of $SG$ and $SE$.

\begin{figure}[!htb]
    \centering
    \begin{subfigure}[b]{0.99\textwidth}
        \includegraphics[width=\textwidth]{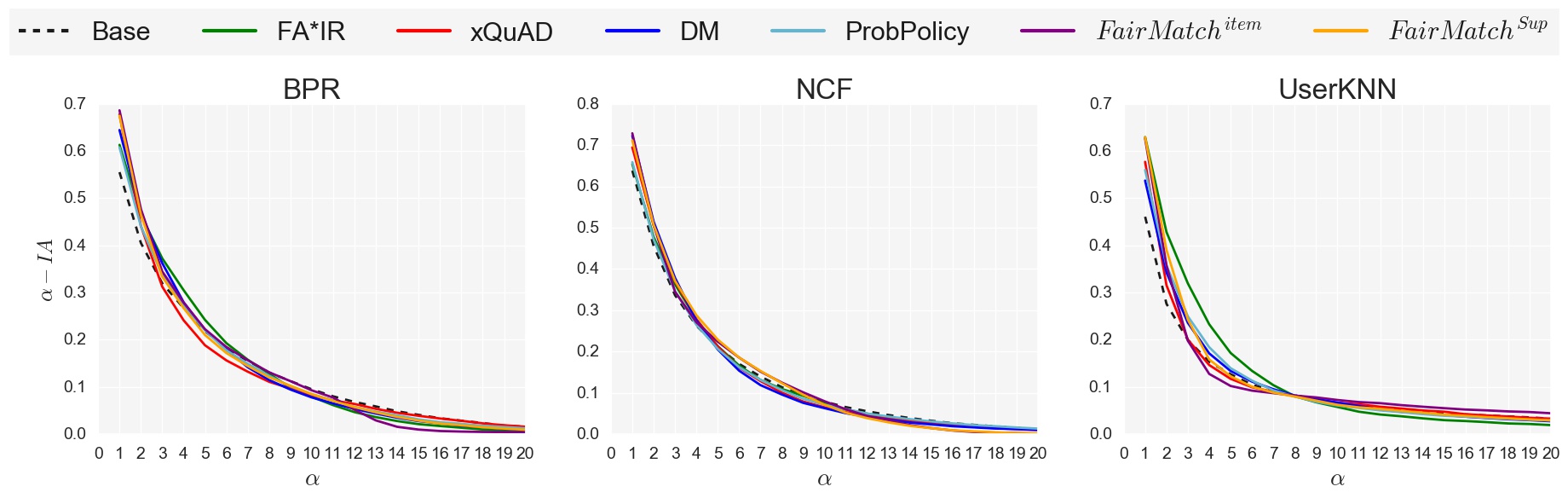}
        \caption{Last.fm} \label{k_a_lf}
    \end{subfigure}
    \begin{subfigure}[b]{0.98\textwidth}
        \includegraphics[width=\textwidth]{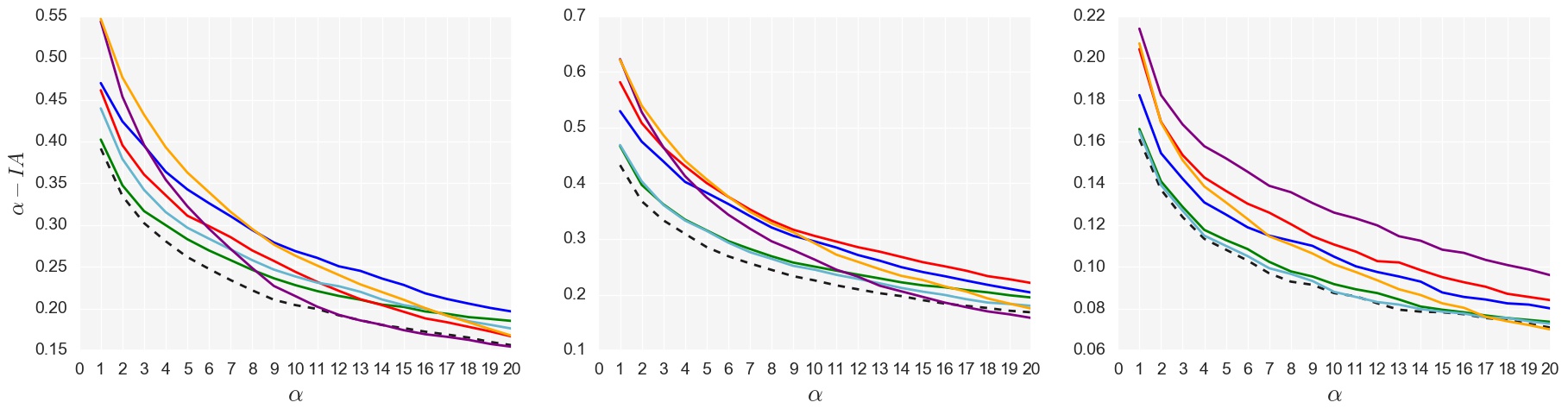}
        \caption{MovieLens} \label{}
    \end{subfigure}%
\caption{Comparison of reranking algorithms in terms of item aggregate diversity ($\alpha\mbox{-}IA$) with different $\alpha$ values.} \label{k_a}
\end{figure}

\subsection{The effect of $\alpha$ in Aggregate Diversity}\label{alphaanalysis}

Standard aggregate diversity metric as it is used in \cite{vargas2011rank,adomavicius2011maximizing} counts an item even if it is recommended only once. Therefore, it is possible for an algorithm to perform really well on this metric while it has not really given enough visibility to different items. For this reason, we introduced $\alpha\mbox{-}IA$ and $\alpha\mbox{-}SA$ which are the generalization of standard aggregate diversity where we only count an item or supplier if it is recommended at least $\alpha$ times. 

\begin{figure}[!htb]
    \centering
    \begin{subfigure}[b]{0.99\textwidth}
        \includegraphics[width=\textwidth]{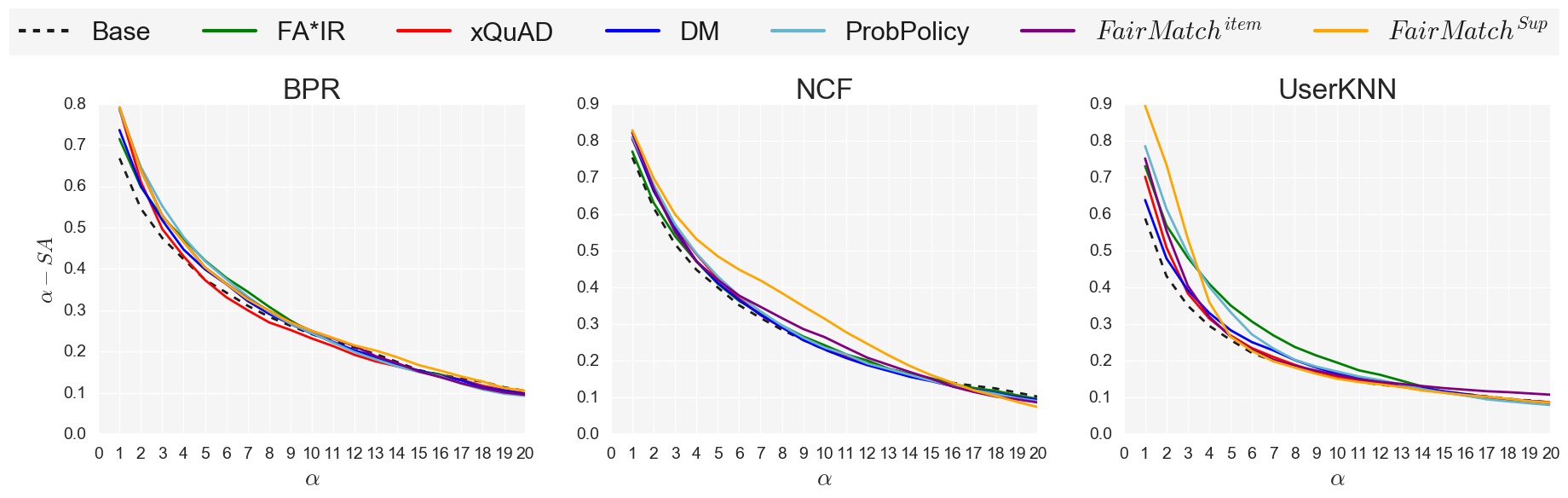}
        \caption{Last.fm} \label{k_sa_lf}
    \end{subfigure}
    \begin{subfigure}[b]{0.98\textwidth}
        \includegraphics[width=\textwidth]{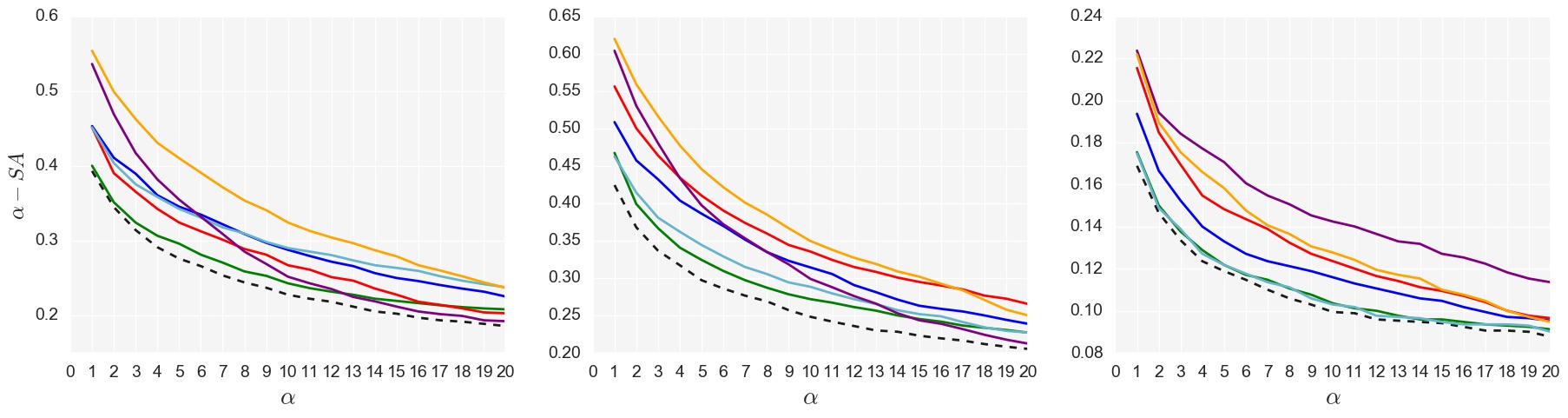}
        \caption{MovieLens} \label{k_sa_ml}
    \end{subfigure}%
\caption{Comparison of reranking algorithms in terms of supplier aggregate diversity ($\alpha\mbox{-}SA$) with different $\alpha$ values.} \label{k_sa}
\end{figure}

Figure \ref{k_a} and \ref{k_sa} show the behavior of different reranking algorithms on aggregate diversity for different values of $\alpha$. The most important takeaway from this figure is that some algorithms perform better than others for smaller values of $\alpha$ while they are outperformed for larger values of $\alpha$. That means if we only look at standard aggregate diversity ($1\mbox{-}IA$ or $1\mbox{-}SA$) we might think a certain algorithm is giving more visibility to different items while in reality that is not the case. For example, using \algname{BPR} as base on MovieLens, $FairMatch^{Sup}$ has better aggregate diversity for smaller values of $\alpha$ ($\alpha \leq8$) than DM while for lager values of $\alpha$ its curve goes under DM indicating lower aggregate diversity. That shows that if we want to make sure different items are recommended more than 8 times, DM would be a better choice but if we want more items to be recommended even if they are recommended less than 8 times, then $FairMatch^{Sup}$ can be better. On supplier aggregate diversity, however, we can see that $FairMatch^{Sup}$ performs better than DM for all values of $\alpha$ indicating no matter how frequent we want the recommended items to appear in the recommendations, $FairMatch^{Sup}$ is still superior.  

\begin{figure*}[btp]
    \centering
    \begin{subfigure}[b]{0.8\textwidth}
        \includegraphics[width=\textwidth]{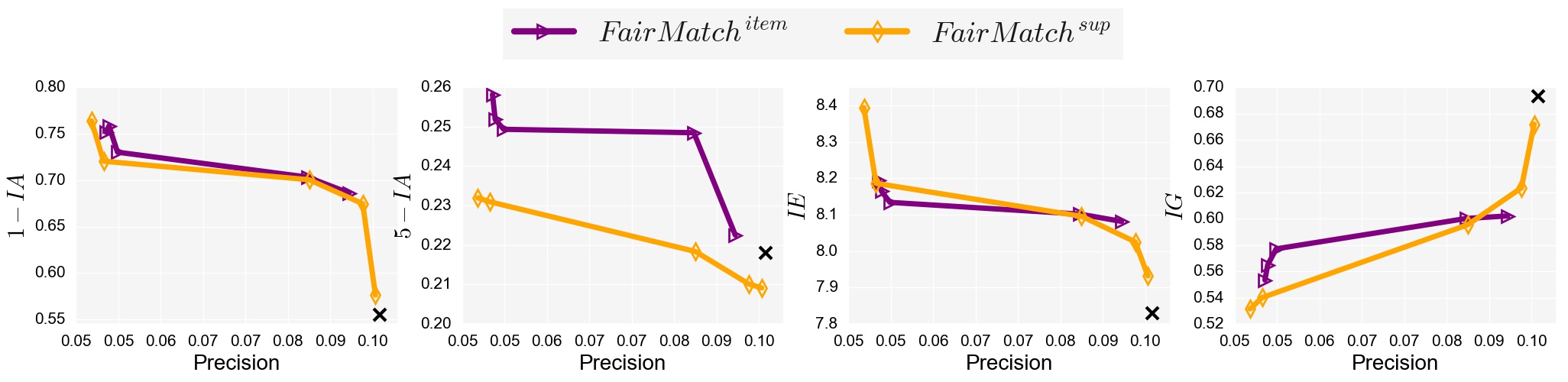}
        \caption{Last.fm, BPR} \label{sensitivity_lf_item_bpr}
    \end{subfigure}
    \begin{subfigure}[b]{0.8\textwidth}
        \includegraphics[width=\textwidth]{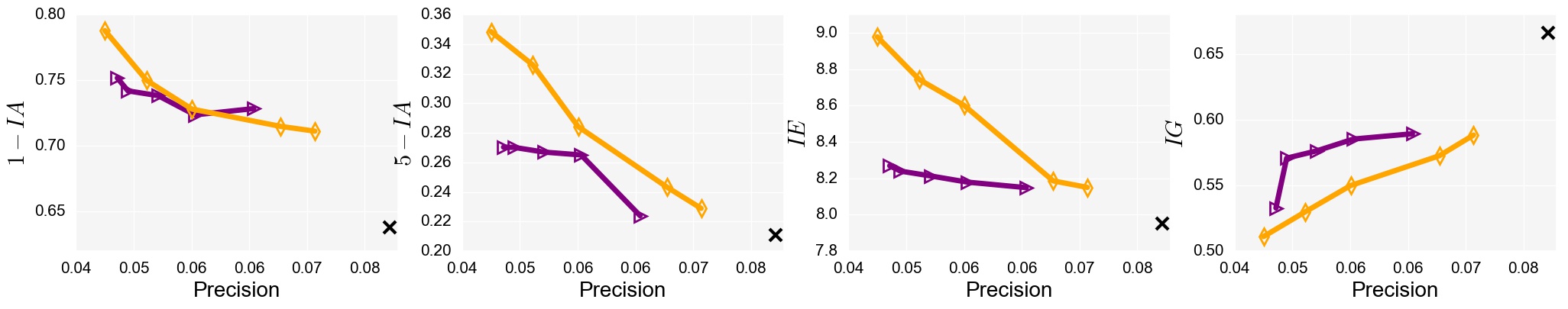}
        \caption{Last.fm, NCF} \label{sensitivity_lf_item_ncf}
    \end{subfigure}
    \begin{subfigure}[b]{0.8\textwidth}
        \includegraphics[width=\textwidth]{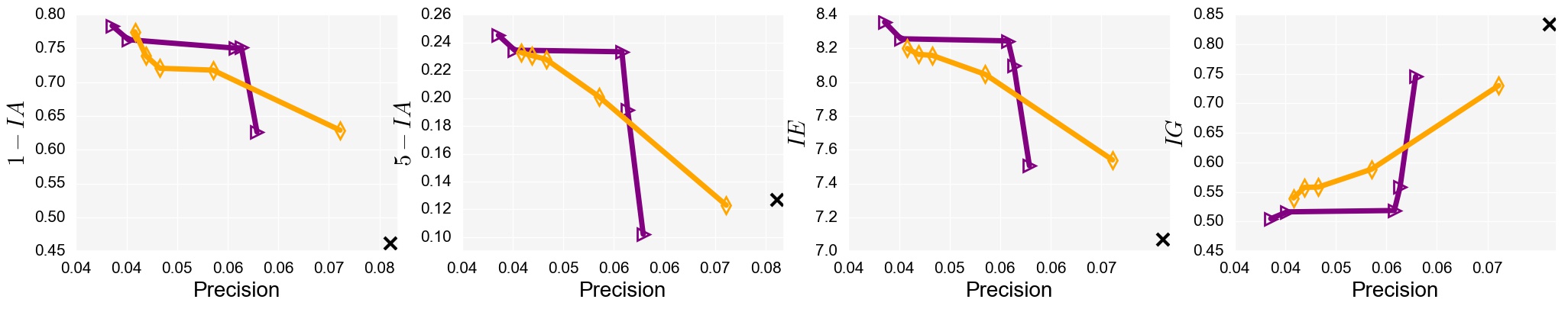}
        \caption{Last.fm, UserKNN} \label{sensitivity_lf_item_userknn}
    \end{subfigure}
    \begin{subfigure}[b]{0.8\textwidth}
        \includegraphics[width=\textwidth]{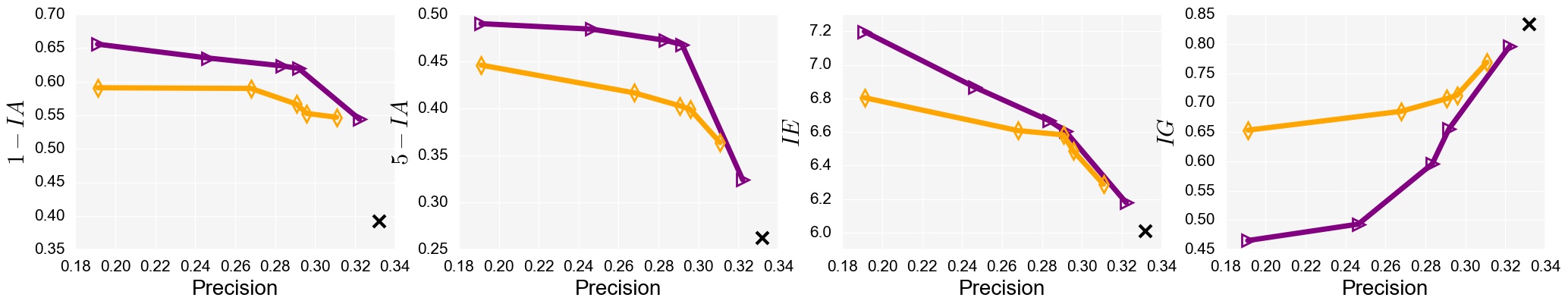}
        \caption{MovieLens, BPR} \label{sensitivity_ml_item_bpr}
    \end{subfigure}
    \begin{subfigure}[b]{0.8\textwidth}
        \includegraphics[width=\textwidth]{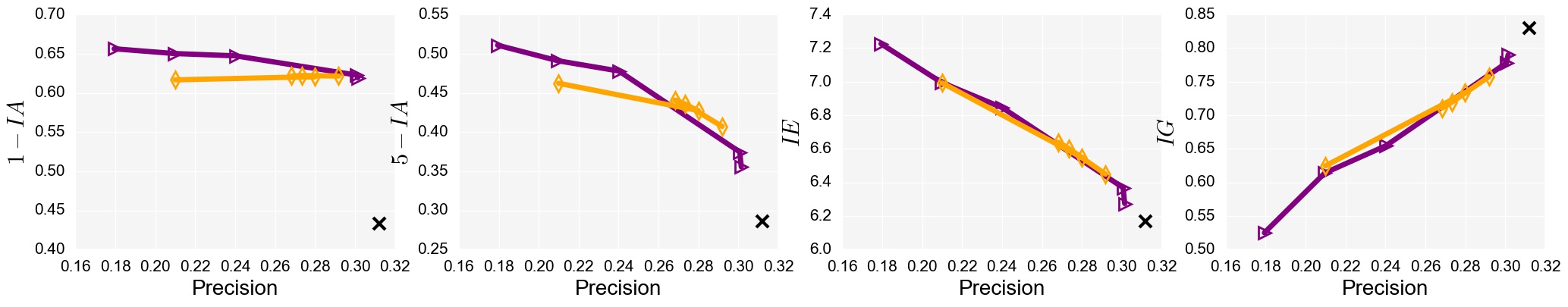}
        \caption{MovieLens, NCF} \label{sensitivity_ml_item_ncf}
    \end{subfigure}
    \begin{subfigure}[b]{0.8\textwidth}
        \includegraphics[width=\textwidth]{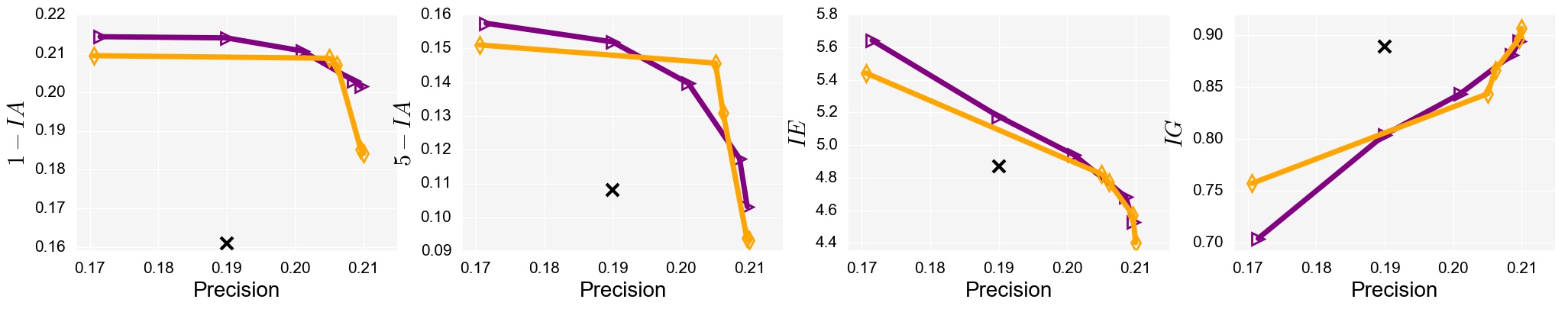}
        \caption{MovieLens, UserKNN} \label{sensitivity_ml_item_userknn}
    \end{subfigure}%
\caption{Trade-off between accuracy and non-accuracy metrics for measuring the exposure fairness of items in FairMatch algorithms on \textbf{Last.fm} and \textbf{MovieLens} datasets using all three base recommenders. The black cross shows the performance of original recommendation lists at size 10.
} \label{sensitivity_item}
\end{figure*}

\subsection{Trade-off between accuracy and non-accuracy metrics for FairMatch}

\begin{figure*}[btp]
    \centering
    \begin{subfigure}[b]{0.8\textwidth}
        \includegraphics[width=\textwidth]{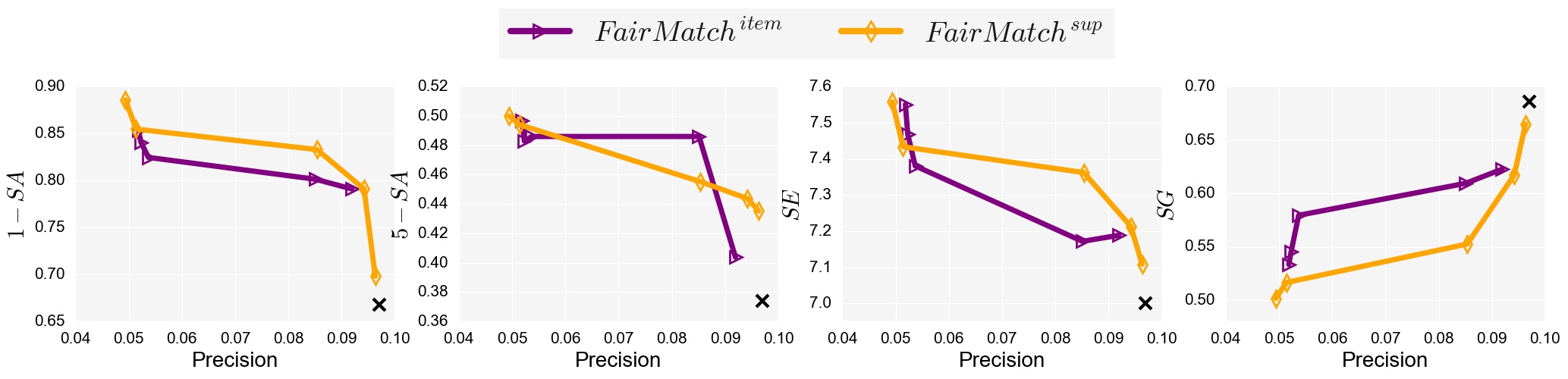}
        \caption{Last.fm, BPR} \label{sensitivity_lf_item_bpr}
    \end{subfigure}
    \begin{subfigure}[b]{0.8\textwidth}
        \includegraphics[width=\textwidth]{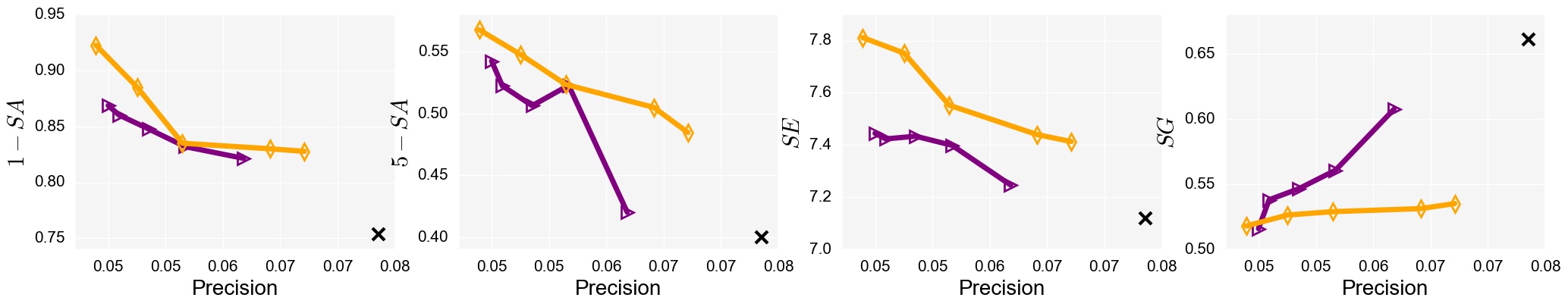}
        \caption{Last.fm, NCF} \label{sensitivity_lf_item_ncf}
    \end{subfigure}
    \begin{subfigure}[b]{0.8\textwidth}
        \includegraphics[width=\textwidth]{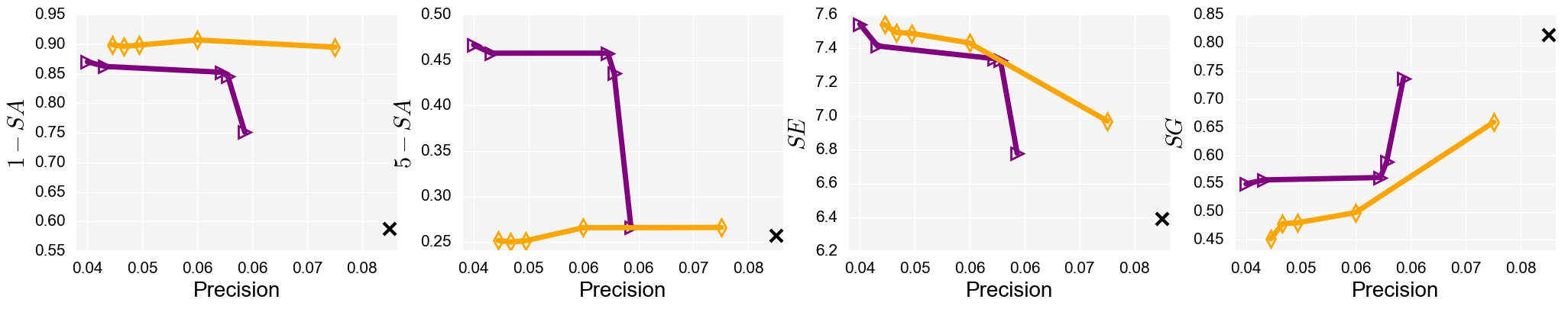}
        \caption{Last.fm, UserKNN} \label{sensitivity_lf_item_userknn}
    \end{subfigure}
    \begin{subfigure}[b]{0.8\textwidth}
        \includegraphics[width=\textwidth]{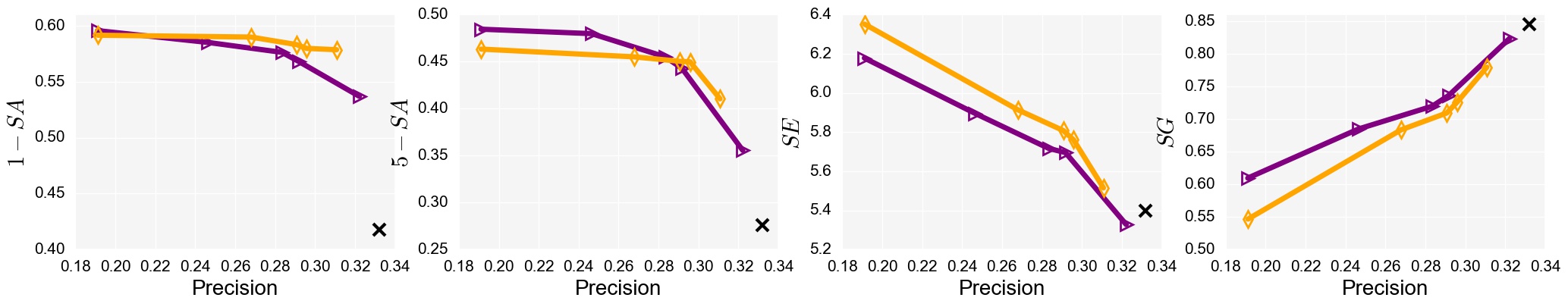}
        \caption{MovieLens, BPR} \label{sensitivity_ml_item_bpr}
    \end{subfigure}
    \begin{subfigure}[b]{0.8\textwidth}
        \includegraphics[width=\textwidth]{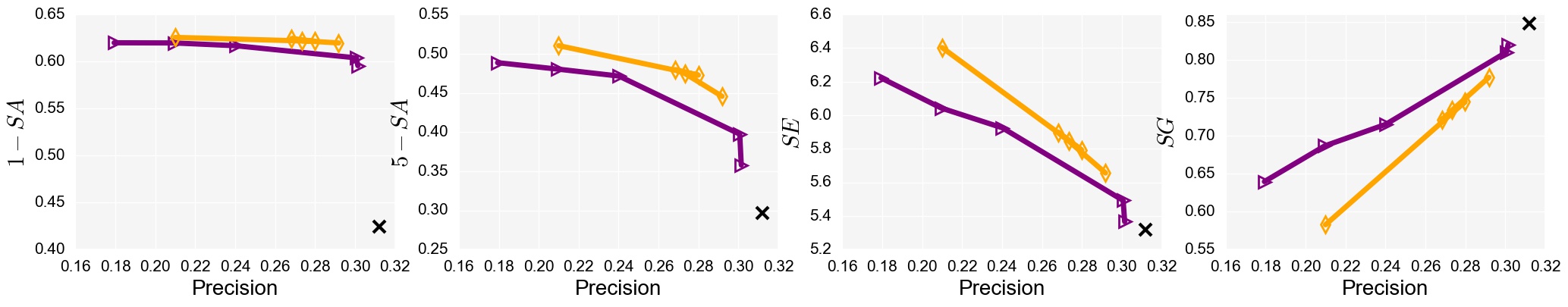}
        \caption{MovieLens, NCF} \label{sensitivity_ml_item_ncf}
    \end{subfigure}
    \begin{subfigure}[b]{0.8\textwidth}
        \includegraphics[width=\textwidth]{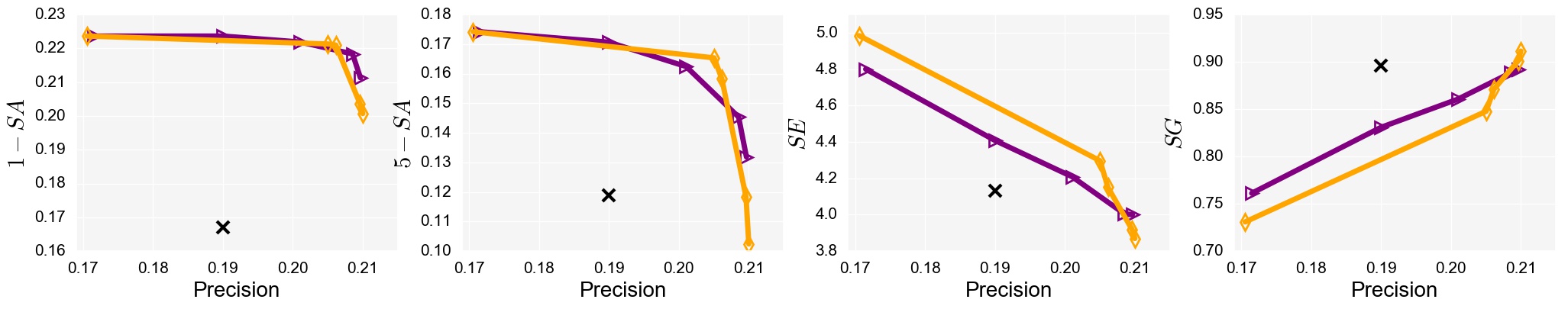}
        \caption{MovieLens, UserKNN} \label{sensitivity_ml_item_userknn}
    \end{subfigure}%
\caption{Trade-off between accuracy and non-accuracy metrics for measuring the exposure fairness of suppliers in FairMatch algorithms on \textbf{Last.fm} and \textbf{MovieLens} datasets using all three base recommenders. The black cross shows the performance of original recommendation lists at size 10.
} \label{sensitivity_supplier}
\end{figure*}

We investigated the trade-off between the precision and non-accuracy metrics under various settings. Figures \ref{sensitivity_item} amd \ref{sensitivity_supplier} show the experimental results for item and supplier exposure, respectively, on Last.fm and MovieLens datasets using all three base recommenders. In these plots, x-axis shows the precision and y-axis shows the non-accuracy metrics (i.e. $1\mbox{-}IA$, $5\mbox{-}IA$, $IE$, and $IG$ in Figure \ref{sensitivity_item} for measuring item exposure and $1\mbox{-}SA$, $5\mbox{-}SA$, $SE$, and $SG$ in Figure \ref{sensitivity_supplier} for measuring supplier exposure) of the recommendation results at size 10. Each point on the plot corresponds to a specific $\lambda$ value and the black cross shows the performance of original recommendation lists at size 10. 

Results in Figure \ref{sensitivity_item} and \ref{sensitivity_supplier} show that $\lambda$ plays an important role in controlling the trade-off between the relevance of the recommended items for users (precision) and improving the utility for items and suppliers (non-accuracy metrics). As we decrease the $\lambda$ value, precision decreases, while non-accuracy metrics increase. According to Equations \ref{wi} and \ref{ws}, for a higher $\lambda$ value, FairMatch will concentrate more on improving the accuracy of the recommendations, while for lower $\lambda$ value, it will have a higher concentration on improving the utility for items and suppliers.

We only report the results for $t=50$, but our analysis on longer initial recommendation lists (e.g. $t=100$) showed that by increasing the size of the initial recommendation lists we will obtain higher improvement on non-accuracy metrics especially on aggregate diversity metrics. However, we will lose accuracy as more items with lower relevance might be added to the final recommendation lists. These parameters allow system designers to better control the trade-off between the precision and non-accuracy metrics.

\subsection{Complexity analysis of FairMatch algorithm}

Solving the maximum flow problem is the core computation part of the FairMatch algorithm. We used Push-relabel algorithm as one of the efficient algorithms for solving the maximum flow problem. This algorithm has a polynomial time complexity as $O(V^{2}E)$ where $V$ is the number of nodes and $E$ is the number of edges in bipartite graph. For other parts of the FairMatch algorithm, the time complexity would be in the order of the number of edges as it mainly iterates over the edges in the bipartite graph.  

Since FairMatch is an iterative process, unlike other maximum flow based techniques \cite{adomavicius2011maximizing, antikacioglu2017}, it requires solving maximum flow problem on the graph multiple times and this could be one limitation of our work. However, except for the first iteration that FairMatch executes on the original graph, at the next iterations, the graph will be shrunk as FairMatch removes some parts of the graph at each iteration. Regardless, the upper-bound for the complexity of FairMatch will be $O(V^{3}E)$ assuming in each iteration we still have the entire graph (which is not the case). Therefore, the complexity of FairMatch is certainly less than $O(V^{3}E)$ which is still polynomial. 

Although FairMatch has a polynomial time complexity, it can be viewed as a limitation for FairMatch as its complexity is still worse that other baselines introduced in this paper. We plan to improve the efficiency of FairMatch algorithm by proposing a unified model in our future work.

\section{Conclusion and future work}

In this paper, we proposed a graph-based approach, FairMatch, for improving the aggregate diversity and exposure fairness of items and suppliers in recommender systems. FairMatch is a post-processing technique that works on the top of any recommendation algorithm. In other words, it reranks the output from the base recommendation algorithms such that it improves the exposure fairness of final recommendation lists with minimum loss in accuracy of recommendations. Experimental results on two publicly available datasets showed that the FairMatch algorithm outperforms several state-of-the-art methods in improving exposure fairness. In addition, we observed that some of the existing metrics for evaluating the performance of recommendation algorithms in terms of popularity mitigation such as aggregate diversity hides important information about the exposure fairness of items and suppliers since this metric does not take into account how frequent different items are recommended. Although Gini can be used to address this issue, it also has its own limitations where an algorithm can achieve a good Gini by equally recommending large number of items or suppliers (even if they are popular) while the rest of items or suppliers still get unfair exposure. Our analysis showed that it is crucial to evaluate bias mitigation algorithms using multiple metrics each of which captures a certain aspect of the algorithm's performance. 
Our definition of exposure fairness in this paper was solely based on the visibility of the items or suppliers in the recommendations without taking into account their original popularity in training data. One possible future work is to take this information into account such that the fairness of exposure for items or suppliers is measured relative to their original popularity as it is done by authors in \cite{abdollahpouri2020addressing}. We also intend to study the effect of our reranking algorithm and other rerankers on the exposure bias of items and suppliers over time where the feedback from the users on recommended items are used in subsequent training step of the recommendation model known as feedback loop.   

Consistent with many prior work on reranking methods, we observed a drop in precision for different rerankers in our offline evaluation setting. However, how users will perceive the recommendations in an online setting can better assess the effectiveness of this type of rerankers. The reason is, the data is skewed towards popular items and it is less likely to observe a hit when recommending less popular items using offline evaluation. Another potential future work is to investigate how users will react to the reranked recommendations by conducting online experiments on real users. 

Finally, in this paper, we studied the ability of FairMatch for improving exposure fairness items and suppliers in recommender systems. However, FairMatch can be generalized to other definitions of fairness
including user fairness. Considering the job recommendation domain
where the task is recommending jobs to users, FairMatch can be
formulated to fairly distribute "good" jobs (e.g. highly-paying jobs)
to each group of users based on sensitive attributes (e.g. men and
women). We consider these scenarios in our future work.

\bibliographystyle{ACM-Reference-Format}
\bibliography{ref}

\end{document}